\documentclass[reprint,superscriptaddress
]{revtex4-1}
\usepackage{graphicx,  color, bm, amsmath, amssymb,hyperref,braket, lipsum, textcomp, mathtools}  
\usepackage[dvipsnames]{xcolor}
\usepackage[normalem]{ulem}
\usepackage[english]{babel}
\usepackage{subfigure}

\DeclareMathOperator{\arccot}{arccot}
\newcommand{\BE}[0]{\begin{equation}}
\newcommand{\EE}[0]{\end{equation}}
\begin{document}
	\title{Probing Majorana bound states via a $pn$ junction containing a quantum dot} 
	\author{L. Bittermann}
	\affiliation{Institut f\"ur Mathematische Physik, Technische Universit\"at Braunschweig, D-38106 Braunschweig, Germany}
	\author{C. \surname{De Beule}}
	\affiliation{Department of Physics and Materials Science, University of Luxembourg, L-1511 Luxembourg, Luxembourg}
	\affiliation{Institut f\"ur Mathematische Physik, Technische Universit\"at Braunschweig, D-38106 Braunschweig, Germany}
	\author{D. Frombach}
	\affiliation{Institut f\"ur Mathematische Physik, Technische Universit\"at Braunschweig, D-38106 Braunschweig, Germany}
	\author{P. Recher}
	\affiliation{Institut f\"ur Mathematische Physik, Technische Universit\"at Braunschweig, D-38106 Braunschweig, Germany}
	\affiliation{Laboratory for Emerging Nanometrology Braunschweig, D-38106 Braunschweig, Germany}
	
	\date{\today}
\begin{abstract}

We propose an alternative route to transport experiments for detecting Majorana bound states (MBSs) by combining topological superconductivity with quantum optics in a superconducting $pn$ junction containing a quantum dot (QD). We consider a topological superconductor (TSC) hosting two Majorana bound states at its boundary ($n$ side). Within an effective low-energy model, the MBSs are coherently tunnel-coupled to a spin-split electron level on the QD which is placed close to one of the MBSs.
Holes on the QD are tunnel-coupled to a normal conducting reservoir ($p$ side). Via electron-hole recombination, photons in the optical range are emitted, which have direct information on the MBS-properties through the recombined electrons. Using a master equation approach, we calculate the polarization-resolved photon emission intensities (PEIs). In the weak coupling regime between MBSs and QD, we find an analytical expression for the PEI which allows to clearly distinguish the cases of well separated MBSs at zero energy from overlapping MBSs. For separated MBSs, the Majorana spinor-polarization is given by the relative widths of the two PEI peaks associated with the two spin states on the QD. For overlapping MBSs, a coupling to the distant (nonlocal) MBS causes a shift of the emission peaks. Additionally, we show that quasiparticle poisoning (QP) influences the PEI drastically and changes its shot noise from super-Poissonian to sub-Poissonian. In the strong coupling regime, more resonances emerge in the PEI due to spin-mixing effects. Finally, we comment on how our proposal could be implemented using a Majorana nanowire.

	\end{abstract}
	
	\maketitle
	
	\section{Introduction}
	Majorana zero modes are quasiparticle states that appear at zero energy in topological superconductors (TSC) at boundaries or in vortices~\cite{Read2000,Kitaev2001,Fu2008} and can be potentially used to build topologically protected qubits~\cite{Kitaev_2003,Nayak2008,beenakker2019}.
	Inspired by the seminal proposal of Fu and Kane~\cite{Fu2008} to combine conventional s-wave superconductors (SC) with Dirac surface states of topological insulators a plethora of hybrid systems of ordinary SCs and appropriate normal systems have been proposed~\cite{Alicea2012,Leijnse2012,Beenakker2013,Aguado2017}. The existence of Majorana zero modes in such hybrid structures have so far been investigated mainly in electron transport setups via tunneling experiments \cite{Mourik2012,Rokhinson2012,Deng2012,Das2012,Churchill2013,Finck2013,Albrecht2016,Deng2016,Chen2017,Suominen2017,Nichele2017,Guel2018,Sestoft2018,Deng2018,Laroche2019} and in Josephson junctions \cite{Kitaev2001,Kwon2004,Fu2009b,San-Jose2012,Dominguez2012,Beenakker2013,Virtanen2013,Houzet2013,Crepin2014,Lee2014,Kane2015,Peng2016,Kuzmanovski2016,Pico-Cortes2017,Dominguez2017,Klees2017,Cayao2017,Sticlet2018,Frombach2020}. Another proposed route is to couple Majorana zero modes to microwave electromagnetic radiation in optical cavities  \cite{OhmHassler,Schmidt,Trif2012,Schmidt_2013,Cottet2013,Ohm2015,Dmytruk2015,Dartiailh2017}, where recently also schemes for braiding and read-out have been proposed as well~\cite{Contamin2021,Trif2019a,Trif2019b}.

\begin{figure}[t]
	\begin{center}
		\includegraphics[width= \columnwidth]{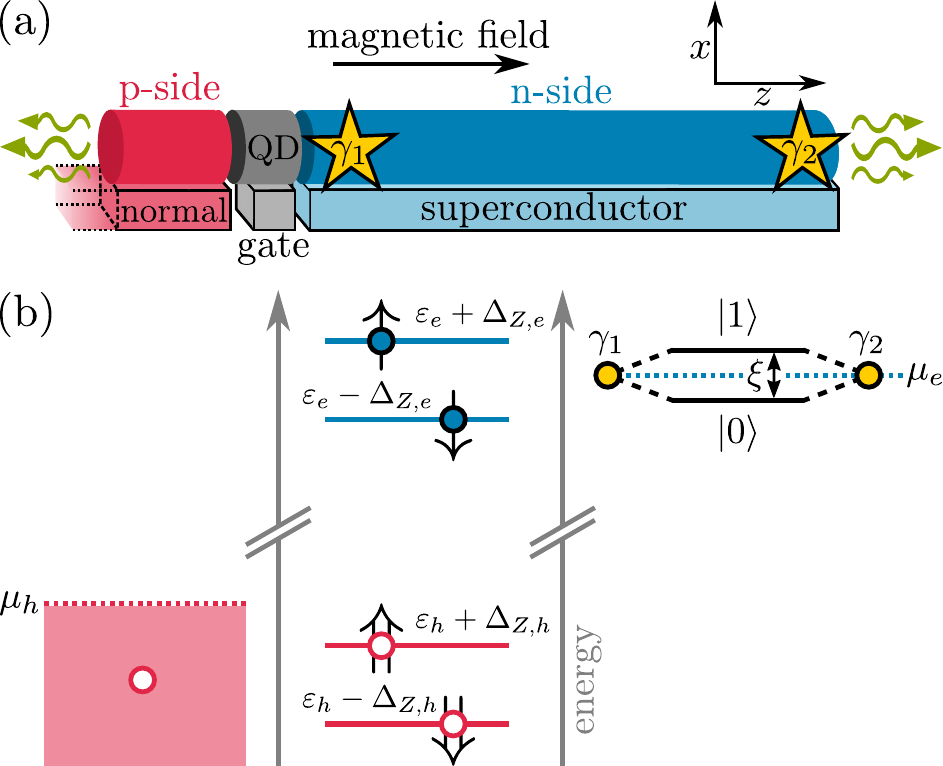} 
		\caption{(a) Exemplary sketch of a $pn$ junction based on semiconducting nanowires in the presence of a magnetic field in $z$-direction with three separated regions. In the topologically non-trivial phase MBSs (yellow stars) emerge at the ends of the $n$ side (blue) which is in proximity to an s-wave SC. The $p$ side is coupled to a normal lead and serves as a hole reservoir (red). In the $pn$ junction a QD is formed (gray) which confines electrons and holes and can be tuned via a gate voltage (gate). Due to electron-hole recombination, photons (green arrows) are emitted in wire direction. (b) Sketch of the energy diagram. Shown are the uncoupled levels of the optically active QD, $n$ side lead (right) and $p$ side lead (left). Full blue and empty red circles represent electrons and heavy holes, respectively.}
		\label{fig:combi_setup} 
	\end{center}
\end{figure}

	Here, we propose another route that has so far been less explored, namely to use an interface between Majorana bound states (MBSs) and an optically active quantum dot (QD). Unlike recently suggested spin-insensitive dipole-dipole coupling of driven excitons coupled to MBSs \cite{Chen2015,Chen2018,Chen2020} our effects rely on a tunnel-coupled spin-split QD in the absence of external electric driving fields. We note that another related idea, considers the direct coupling of two Majorana nanowires forming a light-emitting fractional Josephson junction without access to the spinor wave functions \cite{OhmHassler}. Our setup consists of a $pn$ junction containing an optically active QD, where the $n$ side consists of a TSC and the $p$ side consists of a normal reservoir providing holes to the QD, see Fig.~\ref{fig:combi_setup}. Due to optical selection rules, the polarization of emitted photons is connected to the spin of the QD electron and hole taking part in the recombination process. Since the electronic states of the QD are coherently coupled to the TSC, the signatures of Majorana bound states are imprinted on the emitted photons (via electron-hole recombination) in terms of their energy and polarization. 
	
In particular, we analyze a general model of two MBSs tunnel-coupled to spin-split electron levels on the QD. The electrons are also coherently coupled to heavy-hole states via spontaneous emission of photons. The holes are tunnel-coupled to a normal conducting p-doped lead acting as a reservoir for holes.
We analyze the intensity and noise of the emitted light by using a master equation approach. We identify a regime of interest where two holes are present in the steady state (fast hole refilling), thereby providing a parameter regime where the emitted photons have direct information on the MBSs, their nonlocality \cite{Clarke2017,Prada2017,Schuray2017,Deng2018}, and spinor-polarization \cite{Sticlet2012,Prada2017,Hoffman2017,Schuray2018}, a property not usually available with microwave photons. We provide analytical formulas for the polarization-resolved photon emission intensity (PEI) and its noise in the weak-coupling regime where the spin dynamics becomes decoupled. Information about the mutual Majorana hybridization and the effect of quasiparticle poisoning (QP) and spin-relaxation on the QD are directly observable in our model and results. For larger tunnel couplings between the QD and MBSs and/or larger Majorana splittings, the two spin levels on the QD become effectively coupled by the MBSs which leads to spin-mixing effects in the PEI.

The combination of semiconductor optics and SCs has been a topic of considerable interest in the past few years, including the study of Josephson radiation \cite{JLED}, entangled \cite{Cerletti,Suemune2006,Hassler2010,Nigg2015,Schroer2015} and squeezed \cite{Baireuther2014,Hlobil2015} photons as well as lasing \cite{Godschalk}. There are also experimental realizations of such hybrid systems between SCs and semiconductor optics \cite{SC_LED_exp,Panna2018,Yang2020}. In addition, $pn$ junctions containing optically active QDs were successfully implemented for single \cite{Minot} and entangled photon sources \cite{Versteegh}. We therefore believe that our proposal could be feasible in the realm of quantum wire based setups. Rashba-nanowires in proximity to an s-wave SC and subjected to a magnetic field have been proposed as a platform for MBSs \cite{Lutchyn,Oreg}, and the coherent tunnel-coupling of such wires to a QD has been investigated theoretically \cite{Leijnse2011,Hoffman2016,Prada2017,Chevallier2018,Schuray2020}, as well as experimentally \cite{Deng2016,Deng2018}.

The paper is organized as follows: In Sec. \ref{sec:model}, we introduce the model of a $pn$ junction containing a QD coupled to MBSs and explain how electrons and holes recombine to photons. In Sec. \ref{sec:master_equation}, we discuss the structure of the master equation. Besides the recombination rates for the emission of photons, we consider additional phenomenological rates. We give the polarization-resolved PEI and explain the relevant transitions in the system. In Sec. \ref{sec:results}, we investigate different parameter regimes of the MBSs and present the corresponding PEIs. For decoupled spins we derive analytical expressions for the PEI and compare them to our findings from the full model. Furthermore, we investigate the photon shot noise and the influence of spin relaxation and QP. Additionally, we investigate the spin-mixing regime where electron spins are effectively coupled through the MBSs. In Sec. \ref{sec:nanowire}, we comment on how our proposed $pn$ junction could be implemented using a semiconducting nanowire which can host MBSs. We conclude the paper in Sec. \ref{sec:conclusion}. In the appendices, we show further details of the calculations, in particular the full master equation and the derivation of the photon shot noise.

\section{Model}\label{sec:model}
We consider a $pn$ junction containing a QD with electrons (holes) in the conduction (valence) band with a bias voltage $\mu_n-\mu_p=eV$. The $n$ side is a TSC with chemical potential $\mu_e=\mu_n$ whereas the $p$ side is a normal lead which acts as a hole reservoir with chemical potential $\mu_h=-\mu_p$. In the presence of electron-hole-recombination under photon emission, the model reads 
\begin{align}
H=H_e+H_h+H_{\text{photon}}+V_{\rm rec},
\end{align}
where $H_e$ describes electrons in the conduction band of the QD coherently coupled to the MBSs by tunneling, $H_h$ describes holes in the valence band of the QD, and $H_{\text{photon}}$ gives the photon continuum. Electrons and holes on the QD recombine to photons via $V_{\text{rec}}$.

A possible realization of the setup is shown in Fig.~\ref{fig:combi_setup} where we propose a semiconducting nanowire in a magnetic field (assumed to be in the $z$-direction) with three separated regions: The $n$ side is in proximity to an s-wave SC where the nanowire hosts MBSs in the topologically non-trivial phase, whereas the $p$ side is coupled to a normal lead serving as a hole reservoir. In the $pn$ junction of the wire a QD is formed where electrons and holes can recombine to photons. We note, however, that our proposal could be implemented with any TSC coupled to a QD, embedded in a $pn$ junction. Our results to be derived in the following would hold accordingly. To be specific, we will later comment on the Majorana nanowire model explicitly in Sec. \ref{sec:nanowire}.

We represent $H_e$ by an effective low-energy model, where we can neglect the ordinary proximity effect on the QD induced by the coupling to the superconducting continuum (see below). Therefore, the TSC is modeled only by the MBSs and their mutual coupling. $H_e$ is then given by
\begin{align}\label{eq:HamEl}
\begin{split}
H_e=&\sum_{\sigma=\uparrow,\downarrow}\varepsilon_{e\sigma} d_{\sigma}^\dagger d_{\sigma} + U_e   \hat{n}_{d\uparrow} \hat{n}_{d\downarrow} \\
&+\frac{i}{2}\xi \gamma_1 \gamma_2 +\sum_{\substack{i=1,2\\
\sigma=\uparrow,\downarrow}}\left(t_{i\sigma} d_\sigma \gamma_i+ \text{H.c.}\right).
\end{split}
\end{align}
The first term in Eq.~\eqref{eq:HamEl} corresponds to the QD where the operator $d_\sigma^{(\dagger)}$ annihilates (creates) an electron on the QD with spin $\sigma = \ \uparrow,\downarrow$ along the $z$-axis (having total angular momentum $j_z=\pm\hbar/2$) and energy $\varepsilon_{e\sigma}=\varepsilon_e+\sigma\Delta_{Z,e}$ where $\Delta_{Z,e}$ is the Zeeman energy on the QD.
We count energies from the chemical potential $\mu_e$ of the SC (the $n$ side reservoir).
The second term describes the energy penalty for double occupancy on the QD with charging energy $U_e$ and the occupation number operator $\hat{n}_{d\sigma}=d_\sigma^\dagger d_\sigma $. The third term describes the MBSs $\gamma_1$ and $\gamma_2$ with splitting energy $\xi$ due to a finite overlap of the Majorana wave functions. The MBSs fulfill the self-adjoint condition $\gamma_i^\dagger=\gamma_i$ and $\{\gamma_i,\gamma_j\}=2\delta_{ij}$ with $i=1,2$. The last term in Eq.~\eqref{eq:HamEl} describes the tunneling of electrons with spin $\sigma$ between the QD and MBSs $\gamma_i$ with tunneling amplitudes $t_{i\sigma}$. Here, it is important to keep in mind that the Majorana wave function is a four component spinor whose components are spin-dependent and lead to tunneling amplitudes $t_{i\uparrow}$ and $t_{i\downarrow}$, respectively. Henceforth, we refer to the spin of the electronic components of the MBSs as the spin of the MBSs \cite{Sticlet2012,Prada2017}. Note that $\gamma_1$ and $\gamma_2$ are spatially separated, so that the total tunneling amplitudes $t_i=\sqrt{|t_{i\uparrow}|^2+|t_{i\downarrow}|^2}$ obey $t_1>t_2$, if the QD is placed closer to $\gamma_1$.
In the basis of the nonlocal fermion $c^\dagger$ comprised by two MBSs $\gamma_1=c^\dagger+c$ and $\gamma_2=i(c^\dagger-c)$, the Hamiltonian reads
\begin{align}\label{eq:H_e}
\begin{split}
H_e&=\sum_{\sigma=\uparrow,\downarrow} \varepsilon_{e\sigma}  d_{\sigma}^\dagger d_{\sigma} + U_e \hat{n}_{d\uparrow} \hat{n}_{d\downarrow} +\xi\left(
\hat{n}_c-\frac{1}{2}\right)\\ &+\sum_{\sigma=\uparrow,\downarrow}\left(t_{P\sigma} d_\sigma c + t_{T\sigma} d_\sigma c^\dagger +\text{H.c.}\right),
\end{split}
\end{align}
where $\hat{n}_c=c^\dagger c$, $n_c=0,1$, is the occupation number operator of the nonlocal fermion, $t_{P\sigma}=t_{1\sigma}-it_{2\sigma}$ is a nonlocal pairing between QD and MBSs and $t_{T\sigma}=t_{1\sigma}+it_{2\sigma}$ describes tunneling processes of single fermions. We want to study the properties of the MBS with photons that are emitted via optical transitions, when electrons on the QD recombine with holes in the valence band. We focus on the regime where the singly occupied QD levels are close to the chemical potential $\mu_e$ of the TSC and assume that $U_e\gg\Delta_{Z,e},|t_{i\sigma}|$ \footnote{We include the charging energy $U_e$ in the numerical calculations and assume the value $U_e=10\Delta_{Z,e}$}, see \cite{Deng2016}.
In this case, the states $\ket{\uparrow\downarrow,n_c}$ are off-resonant (i.e the hybridization with the electron reservoir is suppressed). Hence, we do not further discuss the doubly occupied QD state in the remainder of the paper. We assume in addition that $U_e\gg\Delta_S$, where $\Delta_S$ is the superconducting pairing potential in the bulk SC, so the ordinary proximity effect involving tunneling of Cooper pairs between QD and SC is suppressed. Note that in the topologically trivial regime without MBSs, the PEI would be very weak, since the photon emission would be significant only if $U_e<\Delta_S$. In this regime, the PEIs would have no spin dependence (i.e. the emission of $\uparrow$- and $\downarrow$ electrons would be equally strong), since only Cooper pairs can tunnel onto the QD \cite{JLED}. Due to superconducting terms such as $d_\sigma c$ in Eq.~\eqref{eq:H_e} only the fermion parity $(n_{d\uparrow}+n_{d\downarrow}+n_c)$ mod $ 2$ is conserved and even and odd states are decoupled. 
We show $H_e$ in the product basis $\ket{n_{d\sigma}}\times\ket{n_c}$. In the even parity subspace with basis $\{\ket{0,0},\ket{\uparrow,1},\ket{\downarrow,1}\}$, $H_e$ becomes
\begin{align}\label{eq:H_even}
H_e^{\rm even}=
\left(
\begin{array}{cccc}
-\frac{\xi}{2} & -t_{P\uparrow} & -t_{P\downarrow} \\
-t_{P\uparrow}^* & \varepsilon_e+\Delta_{Z,e}+\frac{\xi}{2} & 0 \\
-t_{P\downarrow}^* &  0 & \varepsilon_e-\Delta_{Z,e}+\frac{\xi}{2}
\end{array}
\right),
\end{align}
and in the odd parity basis $\{\ket{0,1},\ket{\uparrow,0},\ket{\downarrow,0}\}$,
\begin{align}\label{eq:H_odd}
H_e^{\rm odd}=
\left(
\begin{array}{cccc}
+\frac{\xi}{2} & -t_{T\uparrow} & -t_{T\downarrow} \\
-t_{T\uparrow}^{*} & \varepsilon_e+\Delta_{Z,e}-\frac{\xi}{2} & 0 \\
-t_{T\downarrow}^* &  0 & \varepsilon_e-\Delta_{Z,e}-\frac{\xi}{2} 
\end{array}
\right).
\end{align}
If we compare $H_e^{\rm even}$ and $H_e^{\rm odd}$, the diagonals only differ by $\xi$, i.e. the energy to occupy the nonlocal fermion. Since the even subspace describes states of zero and two fermions, unoccupied and singly occupied QD levels are connected via $t_{P\sigma}$, see Eq.~\eqref{eq:H_even}, where pairs of fermions are created and annihilated.
	\begin{figure*}[t!]
	\begin{center}
		\includegraphics[width=\textwidth]{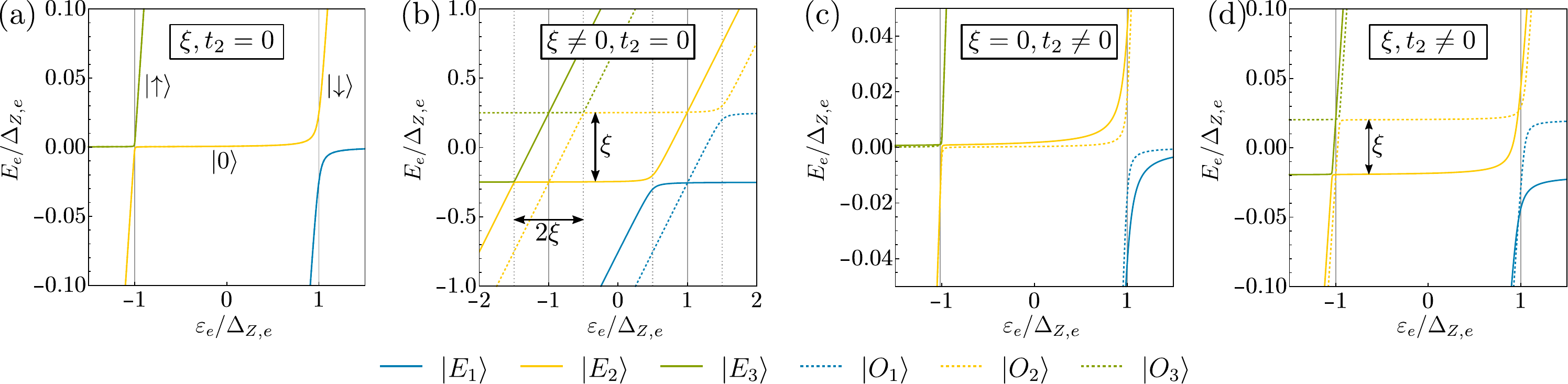}
		\caption{Eigenenergies of the coupled QD-MBSs system over QD energy $\varepsilon_e$ in different regimes. (a) Separated MBSs. Even and odd states are degenerate and we show the QD states $\ket{n_{d\sigma}}$ away from the avoided crossings. Parameters are $\xi=0$, $10t_{1\uparrow}=t_{1\downarrow}=0.025\Delta_{Z,e}$, $t_2=0$. (b) Overlapping MBSs with finite splitting. Parameters are $\xi=0.5\Delta_{Z,e}$, $10t_{1\uparrow}=t_{1\downarrow}=0.05\Delta_{Z,e}$, $t_2=0$. (c) Overlapping MBSs with finite coupling to $\gamma_2$. Parameters are $\xi=0$, $10t_{1\uparrow}=t_{1\downarrow}=0.03\Delta_{Z,e}$, $it_{2\uparrow}=0.0012\Delta_{Z,e}$, $it_{2\downarrow}=-0.012\Delta_{Z,e}$. (d) Overlapping MBSs with finite splitting and finite coupling $t_2$. Parameters are $\xi=0.04\Delta_{Z,e}$, $10t_{1\uparrow}=t_{1\downarrow}=0.03\Delta_{Z,e}$, $it_{2\uparrow}=0.0009\Delta_{Z,e}$, $it_{2\downarrow}=-0.009\Delta_{Z,e}$. }
		\label{fig:spectrum_regimes} 
	\end{center}
\end{figure*}
The odd subspace has only states with one fermion, so unoccupied and singly occupied QD levels are connected via tunneling processes with $t_{T\sigma}$, see Eq.~\eqref{eq:H_odd}. Note that $t_{P\sigma}$ and $t_{T\sigma}$ are different if there is a finite tunneling amplitude $t_2$ to the second MBS.

We diagonalize Eq.~\eqref{eq:H_e} in the product basis $\ket{n_{d\sigma},n_c}$. The resulting eight eigenstates $\ket{\psi}_e=\ket{E_m(O_m)}$ with $m=1,2,3,4$ are sorted by energy and even (odd) parity. Note that the eigenstates $\ket{E_4(O_4)}~\approx~\ket{\uparrow\downarrow,n_c}$ are energetically separated from the other states due to a large charging energy $U_e$ and are discarded in the discussion (see Ref. [83] for further details).
We can identify four qualitatively different parameter regimes depending on the splitting $\xi$ and the coupling $t_2$ to $\gamma_2$. 
In Fig.~\ref{fig:spectrum_regimes}, we show the corresponding spectrum for each regime. In all cases we assume $|t_{1\downarrow}|>|t_{1\uparrow}|$, since this is the case for a finite Zeeman field in a Majorana nanowire \cite{Sticlet2012}, for which the coupling to a QD was already experimentally shown \cite{Deng2016}. Furthermore, we assume that the wave functions of $\gamma_1$ and $\gamma_2$ differ by a relative phase factor of $i$, so that $t_{1\sigma}$ and $it_{2\sigma}$ can be considered to be real \cite{DasSarma2012, Prada2017}. In Fig.~\ref{fig:spectrum_regimes}(a), we show the case of separated MBSs ($\xi,t_2=0$). Here, even and odd states are degenerate. Away from the avoided crossings we can identify the three QD level states in each parity sector. The singly occupied QD levels $\ket{\uparrow}$ and $\ket{\downarrow}$ increase linearly in energy for increasing $\Delta_{Z,e}$ and are split in energy by 2$\Delta_{Z,e}$, whereas the unoccupied level $\ket{0}$ is constantly at zero energy. Due to tunneling between QD and MBSs there are two avoided crossings each at $\varepsilon_e=-\Delta_{Z,e}$ ($\varepsilon_e=+\Delta_{Z,e}$), where $\uparrow$($\downarrow$) electrons are in resonance with the MBSs. Since we chose $|t_{1\downarrow}|>|t_{1\uparrow}|$, the hybridization between $\downarrow$ electrons and MBSs is larger. In Figs.~\ref{fig:spectrum_regimes}(b)-\ref{fig:spectrum_regimes}(d), we show overlapping MBSs, where either $t_2$, $\xi$, or both are finite. In Fig.~\ref{fig:spectrum_regimes}(b), a finite splitting $\xi$ leads to an energy difference between even and odd states. Additionally, the avoided crossings shift away from $\varepsilon_e=\pm \Delta_{Z,e}$ to $\varepsilon_e=\pm \Delta_{Z,e}-\xi$ for even and to $\varepsilon_e=\pm \Delta_{Z,e}+\xi$ for odd states, respectively, since the degeneracy between even and odd states is lifted now. In Fig.~\ref{fig:spectrum_regimes}(c), a finite coupling $t_2$ to the second MBS leads to different hybridizations of even and odd states. In this case $|t_{P\downarrow}|>|t_{T\downarrow}|$, so even states are more hybridized than odd states, which can be seen in the spectrum at $\varepsilon_e=+\Delta_{Z,e}$. In Fig.~\ref{fig:spectrum_regimes}(d), $\xi$ and $t_2$ are both finite. Here, even and odd states are separated in energy by $\xi$ and additionally even and odd states couple differently to the QD. Note that for $t_{1\sigma}=\pm i t_{2\sigma}$ either $t_{P\sigma}=0$ or $t_{T\sigma}=0$. This corresponds to the case of coupling to an ordinary complex fermion, where superconductivity is effectively suppressed.

The holes in the valence band of the QD are given by
\begin{equation}\label{eq:holes}
H_h=\sum_{\sigma=\Uparrow,\Downarrow}\varepsilon_{h\sigma} h_\sigma^\dagger h_\sigma + U_h \hat{n}_{h\Uparrow} \hat{n}_{h\Downarrow},
\end{equation}
with charging energy $U_h$. The operator $h_\sigma^{(\dagger)}$ annihilates (creates) a hole with spin $\sigma$ and energy $\varepsilon_{h\sigma}=\varepsilon_{h}+\sigma\Delta_{Z,h}$ \footnote{Since the holes (electrons) are positively (negatively) charged, a gate acting on electrons and holes simultaneously leads to a shift of their energy levels in the opposite way.}. The holes can in general have a different Zeeman energy $\Delta_{Z,h}$ than the electrons due to a different g-factor. We count energies from the chemical potential $\mu_h$ of the $p$ side reservoir. We assume that the relevant carriers are 'heavy' holes so that $h_{\Uparrow/\Downarrow}^\dagger$ creates a hole with total angular momentum projection $j_{z}=\pm 3\hbar/2$ (see for instance \cite{Lodahl2015}). In the eigenbasis of the number operator $\hat{n}_{h\sigma}=h_\sigma^\dagger h_\sigma$ we obtain four hole states $\ket{\psi}_h=\{\ket{0}_h,\ket{\Uparrow},\ket{\Downarrow},\ket{\Uparrow\Downarrow}\}$ which are eigenstates of $H_h$.
The hole states are separated from the electronic states in the QD by 
a large energy of the order of the gap of the host semiconductor which is approximately given by $\mu_e+\mu_h=eV$. 
The holes merely act as recombination partners for the electrons and do not couple directly to the MBSs, since due to the large separation in energy a coherent coupling between holes and MBSs can be neglected. The holes can be refilled by the normal conducting $p$ side reservoir, and we assume that the hole energies satisfy $0>\varepsilon_h+U_h+|\Delta_{Z,h}|$, so that $\varepsilon_h<0$. In that case, the chemical potential $\mu_h$ is always above the energetically highest hole state, so that the holes can be refilled on the QD with a rate $\Gamma_{h}$ and do not tunnel back to the reservoir. Therefore, the stationary state of the holes without the coupling to the photons described below would be $\ket{\Uparrow\Downarrow} $.

The photon bath is described by
\begin{equation}\label{eq:photons}
H_{\text{photon}}=\sum_k\sum_{\substack{ P=L,R}}\hbar\omega_{k} \ a_{k,P}^\dagger a_{k,P},
\end{equation}
where photons of energy $\hbar\omega_k$ with wave number $k$ and polarization $P$ are annihilated (created) by $a_{k,P}^{(\dagger)}$. The polarization can be $L$-circular with total angular momentum $j_z=-\hbar$ or $R$-circular with $j_z=+\hbar$.

The recombination of electrons and holes in the $pn$ junction containing the QD is given by
\begin{equation}\label{eq:Hrec}
V_{\text{rec}}=g\sum_k\big( d_\uparrow h_\Downarrow a_{k,L}^\dagger + d_\downarrow h_\Uparrow a_{k,R}^\dagger \big) + \text{H.c.}
\end{equation}
with the light-matter-interaction energy $g$, which is a dipole matrix element. It depends on the overlap of the electron and hole wave function and gives rise to selection rules for optical transitions (see for instance \cite{Lodahl2015}). Eq. \eqref{eq:Hrec} describes the emission (absorption) of photons with a given polarization $P$ due to recombination (creation) of an electron and a hole, whose spins are quantized along the magnetic field. 
We assume that this axis corresponds to the high symmetry axis of the QD so that the total angular momentum along this axis commutes with the Hamiltonian, i.e. photons emitted in this direction are circularly polarized. The form of $V_{\text{rec}}$ then holds for emission along this axis (see for instance \cite{SelectionRules}). Photons emitted in other directions would lead to different photon polarizations and $V_{\text{rec}}$ would have to be adjusted accordingly \cite{Gywat,Cerletti}. 

\section{Master equation}\label{sec:master_equation}
The dynamics of the system is governed by electron-hole recombination with the simultaneous emission of a photon changing the states of electrons and holes on the QD. 
Electrons on the QD are coherently coupled to the MBSs forming the eigenstates $\ket{\psi}_e=\ket{E_m(O_m)}$, $m=1,2,3,4$. Incoherent processes are suppressed in the regime $k_BT, |t_{i\sigma}| \ll  \Delta_S$. 
The hole states $\ket{\psi}_h$, however, are incoherently coupled to a normal $p$ side reservoir which we incorporate by rates $\Gamma_h$ (we assume $-(\varepsilon_h+U_h+|\Delta_{Z,h}|)\gg \hbar\Gamma_h$ to have incoherent transitions) which provide new holes to the QD after each photon-emission process. 

We split the Hamiltonian of the open setup into a system part containing the 32 joint electron and hole eigenstates $\ket{\psi}=\ket{\psi}_e\times\ket{\psi}_h$ and treat 
the photons as well as the $p$ side electronic reservoir as a bath. The dynamics of the reduced density matrix of the system can be described by a Markovian master equation, which takes the form of a Pauli master equation \cite{Blum} for the diagonal parts of the reduced density matrix $\langle \psi|\rho(\tau)|\psi\rangle\equiv\rho_{\ket{\psi}}(\tau)$. These elements describe the probability of the system being in the state $\ket{\psi}$ at time $\tau$ and have the following time-evolution
\begin{equation}\label{eq:MasterEq}
	\dot{\rho}_{\ket{\psi}}(\tau)=\sum_{\psi'\neq \psi} \big(-\Gamma_{\ket{\psi'}\leftarrow \ket{\psi}} \rho_{\ket{\psi}}(\tau) +\Gamma_{\ket{\psi}\leftarrow \ket{\psi'}} \rho_{\ket{\psi'}}(\tau)\big),
\end{equation}
where the rates $\Gamma_{\ket{\psi'}\leftarrow \ket{\psi}}$ ($\Gamma_{\ket{\psi}\leftarrow \ket{\psi'}}$) reduce (increase) the occupation $\rho_{\ket{\psi}}(\tau)$ with time $\tau$. The stationary state is given by $\partial_\tau \rho^{\rm stat}_{\ket{\psi}}(\tau)=0.$ Eq.~(\ref{eq:MasterEq}) can also be written as $\dot{\rho}(\tau)={\cal L}\rho(\tau)$ where $\rho(\tau)$ is the vectorized reduced density matrix of its diagonal elements and ${\cal L}$ is the Liouvillian (see App.~\ref{app:fano}). 

In the following paragraphs, we introduce four different kinds of rates that we use in the master equation.
Via optical transitions from an initial state $\ket{\psi_i}$ (energy $E_i$) to a final state $\ket{\psi_f}$ (energy $E_f$) a photon with state $\ket{k,P}=a_{k,P}^\dagger\ket{0}_{\rm ph}$ (energy $\hbar\omega_k$) is emitted. Here, $\ket{0}_{\rm ph}$ is the vacuum state of optical photons.
We only consider photon emission, i.e. there are no optical photons in the initial state. The corresponding recombination rates (see App.~\ref{app:rates}) are calculated with Fermi's Golden Rule
\begin{equation}\label{eq:FGR}
w^P_{\ket{\psi_f} \leftarrow \ket{\psi_i}} (\omega_k)= \frac{2\pi}{\hbar} \big|M_{f,i}^{P}|^2 \delta(E_i-E_f-\hbar\omega_k)
\end{equation}
with matrix elements
\begin{equation}\label{eq:matrixelements}
M_{f,i}^{P}=\bra{\psi_f}\bra{k,P} V_{\text{rec}} \ket{\psi_i} \ket{0}_{\rm ph}.
\end{equation}
The delta function in Eq.~\eqref{eq:FGR} is broadened to a Lorentz curve
$\delta(E_i-E_f-\hbar\omega_k)=d_{\rm ph}/\pi(d_{\rm ph}^2+(E_i-E_f-\hbar\omega_k)^2)$ due to the finite lifetime of electron-hole pairs. 
For simplicity, we assume the same width $d_{\rm ph}$ for all the transitions which is artificially broadened to improve visibility (cf. Figs.~\ref{fig:eff_case1}(e), \ref{fig:eff_case2}(c), and \ref{fig:eff_case2}(d)). Due to energy conservation the energy of emitted photons is given by
$\hbar\omega_k= \mu_e+\mu_h+\Delta E_e + \Delta E_h\sim1$ eV, which is in the optical range.
Here $\Delta E_e$ ($\Delta E_h$) describes the energy difference from an initial to a final state in the subsystem $H_e$ ($H_h$).
The total rate for the transition of the system state from $\ket{\psi_i}$ to $\ket{\psi_f}$ is obtained by integrating Eq.~\eqref{eq:FGR} over $k$. By changing the integration variable from $k$ to $\hbar\omega$ we define $W^P_{\ket{\psi_f} \leftarrow \ket{\psi_i}}=N_{\rm ph}\int d(\hbar\omega) w^P_{\ket{\psi_f} \leftarrow \ket{\psi_i}} (\omega)$ where $N_{\rm ph}$ is the photon density of states assumed to be constant in the frequency range of interest. By integrating out the Lorentz function we obtain the total rate
\begin{equation}
\label{eq:changerate}
W^P_{\ket{\psi_f} \leftarrow \ket{\psi_i}}=\frac{2\pi}{\hbar} |M_{f,i}^{P}|^2 N_{\rm ph},
\end{equation}
which we insert into Eq.~\eqref{eq:MasterEq}.
Note that optical transitions change the parity of the electronic and hole subsystems.
The light-matter-interaction energy $g$ in Eq.~\eqref{eq:Hrec} determines the maximal recombination rate $W_{\rm max}=\frac{2\pi}{\hbar} |g|^2 N_{\rm ph}$.

We supplement the master equation by additional phenomenological rates.
The hole refilling rate $\Gamma_{h}$ refills holes on the QD from a normal reservoir.
It is necessary to have a stationary emission of photons. Indeed for $\Gamma_{h}=0$, the stationary state for the holes would be $\ket{0}_h$ and the total PEI $I_P$ defined below would be zero.
We can add two additional rates that describe environmental influences on the system.
Quasiparticle poisoning changes the occupation of the nonlocal fermion and therefore the parity of the system with rate $\Gamma_{QP}$ \cite{Rainis,Pikulin2012,Cheng2012,San-Jose2012,Virtanen2013,karzig2021}. 
Furthermore, we assume a spin relaxation rate $\Gamma_R$ 
that flips the spin of the energetically higher $\uparrow$ electron on the QD which conserves the parity. Note that this is a non-reversible process since the spin excitation process is not probable due to a large energy difference between spin states ($k_BT\ll\Delta_{Z,e}$) \footnote{We note that incoherent spin-flips are in principle also possible due to the coupling of QD electrons to the SC continuum with the simultaneous emission of a photon (with an energy shifted by $\Delta_S$) \cite{JLED}. We neglect here these weak processes, but mention that the spin-relaxation part of these processes could be included in $\Gamma_R$.}.
These rates can be included in the master equation Eq.~(\ref{eq:MasterEq}) by adding the following Lindblad superoperators \cite{Cheng2012,San-Jose2012,Pikulin2012} on the right-hand side of the equation and taking its expectation value in the system state $\ket{\psi}$,
\begin{equation}\label{eq:Lindblad_hole}
{\cal L}_{h\sigma}[\rho]=\Gamma_{h}\left(h_{\sigma}^{\dagger}\rho h_{\sigma}-\frac{1}{2}\{\rho,h_{\sigma}h_{\sigma}^{\dagger}\}\right),
\end{equation}
\begin{equation}\label{eq:Lindblad_QP}
\begin{split}
{\cal L}_{QP}[\rho]&=\Gamma_{QP}\bigg(c{\cal P}\rho {\cal P}c^{\dagger}+c^{\dagger}{\cal P}\rho {\cal P}c\\
&-\frac{1}{2}\{\rho,{\cal P}c^{\dagger}c{\cal P}\}-\frac{1}{2}\{\rho,{\cal P}cc^{\dagger}{\cal P}\}\bigg),
\end{split}
\end{equation}
\begin{equation}\label{eq:Lindblad_relax}
{\cal L}_{R}[\rho]=\Gamma_{R}\left(d_{\downarrow}^{\dagger}d_{\uparrow}\rho d_{\uparrow}^{\dagger}d_{\downarrow}-\frac{1}{2}\{\rho,d_{\uparrow}^{\dagger}d_{\downarrow}d_{\downarrow}^{\dagger}d_{\uparrow}\}\right),
\end{equation}
where ${\cal P}=(-1)^{{\hat n}_c}$. The full master equation (see App.~\ref{app:master}) for all eigenstates $\ket{\psi}$ results in 32 differential equations like Eq.~\eqref{eq:MasterEq} with four different kinds of rates we introduced above. We define the frequency-dependent PEI as a function of the frequency $\omega$ for photons with polarization $P$ as
\begin{equation}\label{eq:int}
i_P(\omega)= \hbar N_{\rm ph} \sum_{\psi',\psi} w^P_{\ket{\psi'} \leftarrow \ket{\psi}}(\omega) \rho^{\rm stat}_{\ket{\psi}},
\end{equation}
where $\psi'\neq \psi$, and the polarization-resolved total PEI
\begin{equation}\label{eq:total_int}
I_P =\int d\omega i_P (\omega).
\end{equation}
Before we discuss the photon emission spectra, we examine the different processes described by the master equation in more detail.

\begin{figure}[t]
	\begin{center}
		\includegraphics[width=.8\columnwidth]{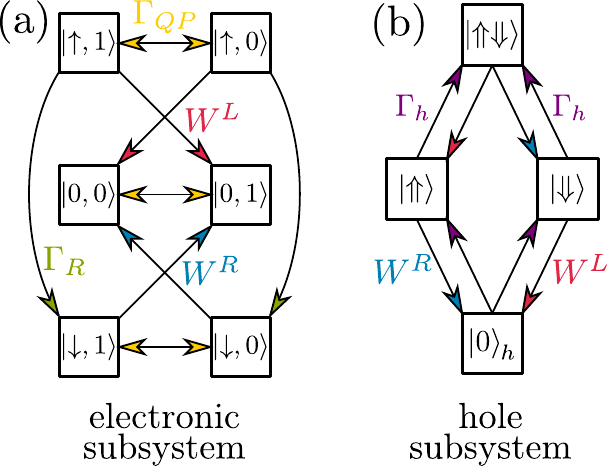}
		\caption{Relevant transitions in electronic and hole subsystems. We show the six relevant uncoupled electronic states in (a) and the four hole eigenstates in (b), respectively. Rates included in the master equation are optical transitions for emission of $L$/$R$-photons with rates $W^L/W^R$ (red/blue), spin relaxation processes with rate $\Gamma_R$ (green), QP with rate $\Gamma_{QP}$ (yellow), and hole refilling with rate $\Gamma_{h}$ (violet).} 
		\label{fig:transitions} 
	\end{center}
\end{figure}

The master equation is written in the eigenstates of the coherently coupled QD-MBSs-system. To better understand the transition processes, it is useful to look at the case where QD and MBSs are decoupled. In Fig.~\ref{fig:transitions}(a), we show the relevant transitions between decoupled states $\ket{n_{d\sigma},n_c}$ of the electronic subsystem.
In particular, we look at the states $\ket{0,0}$, $\ket{\uparrow,1}$, and $\ket{\downarrow,1}$ in the even, and $\ket{0,1}$, $\ket{\uparrow,0}$, and $\ket{\downarrow,0}$ in the odd parity sector, respectively. Optical transitions correspond to $\ket{\uparrow,1}(\ket{\uparrow,0})\xrightarrow{W^L}$~$\ket{0,1}(\ket{0,0})$ or $\ket{\downarrow,1}(\ket{\downarrow,0})\xrightarrow{W^R}$~$\ket{0,1}(\ket{0,0})$ with rates $W^L$ or $W^R$, respectively. In the former case, a $\uparrow$ electron recombines with a $\Downarrow$ hole to a $L$-photon resulting in an empty QD in the electronic subsystem, while in the latter case, $\downarrow$ electrons recombine with $\Uparrow$ holes to a $R$-photon. Note that these processes change the parity in the electronic and hole subsystem, see Fig.~\ref{fig:transitions}(b). Importantly, we use a hole refilling rate $\Gamma_{h}>W_{\text{max}}$, so that holes on average are refilled before the next photon is emitted and are mostly doubly occupied in the stationary state (for an exception, see App.~\ref{app:fano}). Other processes, i.e. spin relaxation and QP, only act in the electronic subsystem, see Fig.~\ref{fig:transitions}(a). Spin relaxation induces transitions $\ket{\uparrow,0}(\ket{\uparrow,1})\xrightarrow{\Gamma_R}\ket{\downarrow,0}(\ket{\downarrow,1})$ with rate $\Gamma_R$ since $\uparrow$ electrons relax to $\downarrow$ electrons preserving the parity. On the other hand, QP with rate $\Gamma_{QP}$ connects states with the same QD occupation but a different occupation of the nonlocal fermion. Therefore, it connects states $\ket{n_{d\sigma},0}\xleftrightarrow{\Gamma_{QP}}\ket{n_{d\sigma},1}$ and changes the parity. When the QD and the MBSs are not coupled, electrons can not be refilled on the QD. Hence, the stationary state of the whole system only contains $\rho^{\rm stat}_{\ket{0,0}\ket{\Uparrow\Downarrow}}$ and $\rho^{\rm stat}_{\ket{0,1}\ket{\Uparrow\Downarrow}}$, such that in this case there is no photon emission in the stationary state.

\section{Results}\label{sec:results}
We now discuss the photon emission of the coupled QD-MBSs-system, where we focus on the signatures of the MBSs and their nonlocality when the QD electrons and MBSs are in resonance. Additionally, we investigate the influence of spin relaxation and QP on the PEI and the photon shot noise. In all cases, we assume $|t_{1\downarrow}|>|t_{1\uparrow}|$, which is the case for the coupling to a Majorana nanowire in a magnetic field under realistic wire parameters \cite{Sticlet2012}. We start with the regime where resonances of $\uparrow$- and $\downarrow$ electrons with the MBSs are well separated. In this case, we can consider them separately with an effectively spinless model. Then we investigate the regime where the resonances are close together which leads to emission processes involving both electron spins.
\subsection{Decoupled spins}
In the case $\Delta_{Z,e}\gg t_1,t_2,\xi$, the $\uparrow$- and $\downarrow$ electrons on the QD remain approximately decoupled when they are coupled to the MBSs. Hence, we can consider the spins separately and use an effectively spinless model
\begin{align}\label{eq:spinless}
\begin{split}
H_\sigma&=\varepsilon_{e\sigma} d_\sigma^\dagger d_\sigma +\xi(\hat{n}_c-\frac{1}{2})\\
&+\big(t_{P\sigma} d_\sigma c + t_{T\sigma} d_\sigma c^\dagger +\rm{H.c.}\big), 
\end{split}
\end{align}
which is Eq.~\eqref{eq:H_e} for a single spin $\sigma= \ \uparrow,\downarrow$. In the product basis $\ket{n_{d\sigma},n_c}$, the four normalized eigenstates are
\begin{subequations}\label{eq:eff_eigenstates}
\begin{align}
\ket{E_{\sigma\pm}}=\frac{(E_{E_{\sigma\mp}}+\frac{\xi}{2})\ket{0,0}+t_{P\sigma}\ket{\sigma,1}}{\sqrt{\left(E_{E_{\sigma\mp}}+\frac{\xi}{2} \right)^2+t_{P\sigma}^2}},\label{eq:eff_eigenstates_even}\\
\ket{O_{\sigma\pm}}=\frac{(E_{O_{\sigma\mp}}-\frac{\xi}{2})\ket{0,1}+t_{T\sigma}\ket{\sigma,0}}{\sqrt{\left(E_{O_{\sigma\mp}}-\frac{\xi}{2} \right)^2+t_{T\sigma}^2}},\label{eq:eff_eigenstates_odd}
\end{align}
\end{subequations}
with eigenenergies
\begin{subequations}
\begin{align}
E_{E_{\sigma\pm}}=\frac{1}{2} \left(\varepsilon_{e\sigma} \pm\sqrt{4 t_{P\sigma}^2+(\varepsilon_{e\sigma}+\xi  )^2}\right),\\
E_{O_{\sigma\pm}}=\frac{1}{2} \left(\varepsilon_{e\sigma} \pm\sqrt{4 t_{T\sigma}^2+(\varepsilon_{e\sigma} -\xi )^2}\right).
\end{align}
\end{subequations}
To solve the rate equation in Eq.~\eqref{eq:MasterEq} analytically, we assume a large hole refilling rate $\Gamma_{h}\to\infty$. This means that after emitting a photon, the hole refilling is so fast that the system always ends up in the doubly occupied hole state $\ket{\Uparrow\Downarrow}$ before the next photon is emitted. In this limit, we only need to include the electronic degrees of freedom in the master equation. Hence, the reduced density matrix is a $4\times4$ matrix with diagonals $\rho_{\ket{E_{\sigma\pm}}}(\tau)$ and $\rho_{\ket{O_{\sigma\pm}}}(\tau)$, where we suppress the hole index. In this case, the rate equation becomes
\begin{subequations}\label{eq:mastereq_spinless}
\begin{align}
\begin{split}
\dot{\rho}_{\ket{E_{\sigma n}}}(\tau)\\
=\sum_{m=\pm}\Big[&-\left(W^P_{\ket{O_{\sigma m}}\leftarrow \ket{E_{\sigma n}}}+\Gamma_{QP}^{\ket{O_{\sigma m}}\leftarrow \ket{E_{\sigma n}}}\right)\rho_{\ket{E_{\sigma n}}}(\tau)\\
&+\left(W^P_{\ket{E_{\sigma n}}\leftarrow \ket{O_{\sigma m}}}+\Gamma_{QP}^{\ket{E_{\sigma n}}\leftarrow \ket{O_{\sigma m}}}\right)\rho_{\ket{O_{\sigma m}}}(\tau)\Big],
\end{split}\\
\begin{split}
\dot{\rho}_{\ket{O_{\sigma n}}}(\tau)\\
=\sum_{m=\pm}\Big[&-\left(W^P_{\ket{E_{\sigma m}}\leftarrow \ket{O_{\sigma n}}}+\Gamma_{QP}^{\ket{E_{\sigma m}}\leftarrow \ket{O_{\sigma n}}}\right)\rho_{\ket{O_{\sigma n}}}(\tau)\\
&+\left(W^P_{\ket{O_{\sigma n}}\leftarrow \ket{E_{\sigma m}}}+\Gamma_{QP}^{\ket{O_{\sigma n}}\leftarrow \ket{E_{\sigma m}}}\right)\rho_{\ket{E_{\sigma m}}}(\tau)\Big],
\end{split}
\end{align}
\end{subequations}
for $n=\pm$. Here, we calculate the recombination rates $W^P_{\ket{\psi}\leftarrow \ket{\psi'}}$ with Eq.~\eqref{eq:changerate} and the QP rates
\begin{align}
\begin{split}
\Gamma_{QP}^{\ket{\psi}\leftarrow \ket{\psi'}}= \Gamma_{QP}(|\bra{\psi}c{\cal P}\ket{\psi'}|^2
+|\bra{\psi}c^\dagger{\cal P}\ket{\psi'}|^2),
\end{split}
\end{align}
according to Eq.~\eqref{eq:Lindblad_QP}.
Note that we do not include spin relaxation processes within the spinless model which would couple the two spin sectors. Without QP ($\Gamma_{QP}=0$), we find that the total PEI is given by
\begin{align}\label{eq:eff_int}
\frac{I_\sigma}{I_P^{\rm max}}=I_0\frac{d^2}{d^2+(\varepsilon_{e\sigma}-\varepsilon_0)^2}
\end{align}
with $I_P^{\rm max}=W_{\rm max}/2$, which has the form of a Lorentzian with
\begin{subequations}\label{eq:Lorentzian_parameters}
\begin{align}
I_0&=\frac{t_{P\sigma}^2+t_{T\sigma}^2}{t_{P\sigma}^2+t_{T\sigma}^2+\xi^2}\label{eq:int_maximum},\\
d^2&=\frac{4t_{P\sigma}^2 t_{T\sigma}^2(t_{P\sigma}^2+t_{T\sigma}^2+\xi^2)}{(t_{P\sigma}^2+t_{T\sigma}^2)^2}\label{eq:int_width},\\
\varepsilon_0&=\frac{t_{P\sigma}^2-t_{T\sigma}^2}{t_{P\sigma}^2+t_{T\sigma}^2}\xi\label{eq:int_shift},
\end{align}
\end{subequations}
where $I_0$, $2d$ and $\varepsilon_0$ correspond to the height, width and location of the peak, respectively. Here, the Lorentzian shape of the PEI reflects the coherent coupling between QD electrons and MBSs. In the following, we will investigate the results of our numerical calculations and compare them to the effectively spinless model.

\subsubsection{Separated MBSs}
\begin{figure}[t!]
	\begin{center}
		\includegraphics[width=1\columnwidth]{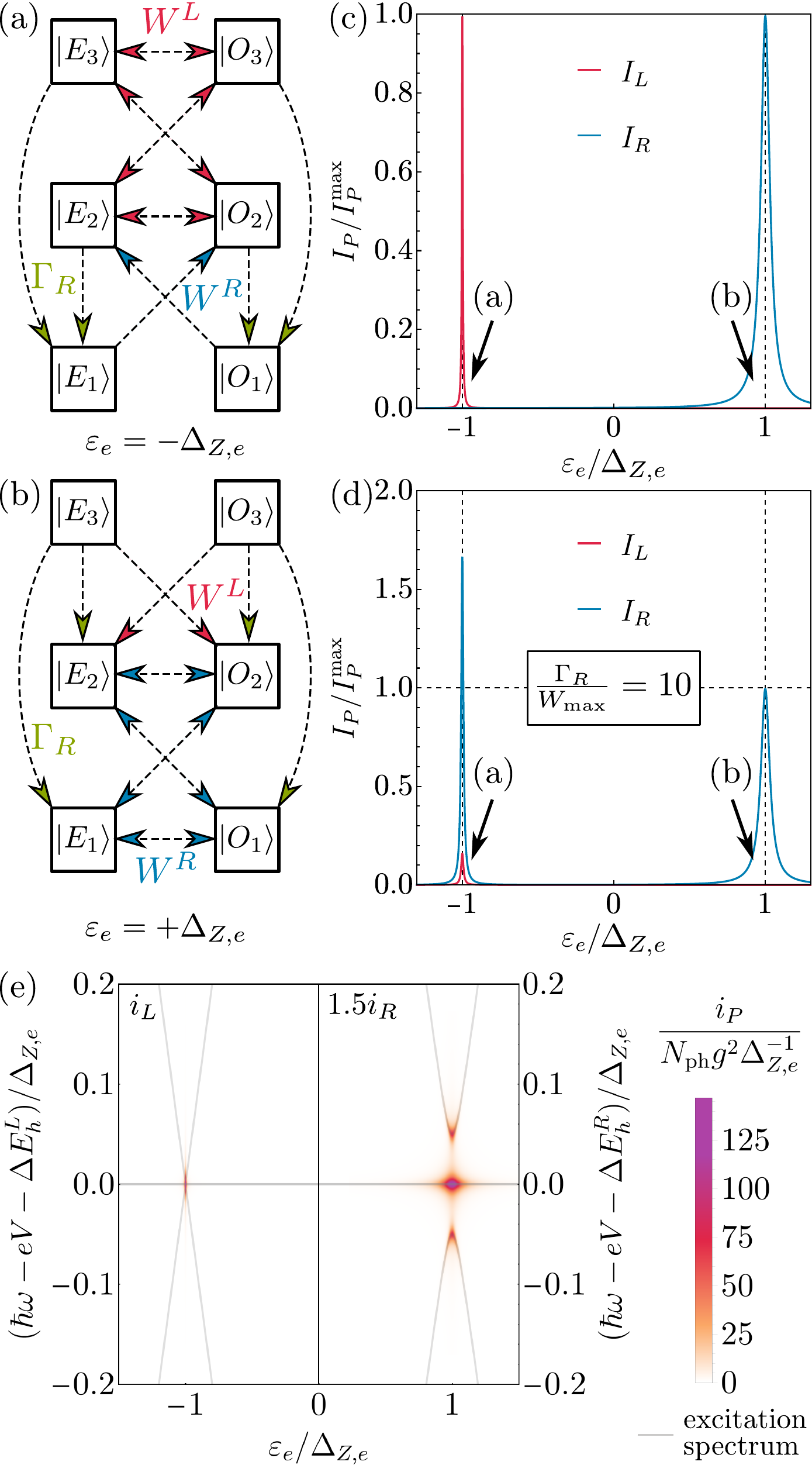}
		\caption{Photon emission for separated MBSs ($\xi=t_2=0$). Transition schemes for $\varepsilon_e=-\Delta_{Z,e}$ (a) and $\varepsilon_e=+\Delta_{Z,e}$ (b). Here, $W^L$ and $W^R$ are recombination rates giving $L$- and $R$-photons, respectively, whereas $\Gamma_R$ is the spin relaxation rate. The total PEI $I_P$ is shown for $\Gamma_R=0$ in (c) and for $\Gamma_R=10W_{\rm max}$ in (d). (e) Frequency-dependent PEI $i_P$ over the QD energy $\varepsilon_e$ and the photon energy $\hbar\omega$ for same parameters as in (c), where $\Delta E_h^{L/R}=U_h+\varepsilon_{h\Downarrow/\Uparrow}$. Shown are only transitions from the doubly occupied hole state. Gray lines correspond to the excitation spectrum of the coupled QD-MBSs-system. Further parameters are $10t_{1\uparrow}=t_{1\downarrow}=0.025\Delta_{Z,e}$, $\Gamma_{h}= 100W_{\text{max}}$, $\Gamma_{QP}=0$.} 
		\label{fig:eff_case1} 
	\end{center}
\end{figure}

First, we consider the case of well separated MBSs with $\xi=t_2=0$. In Fig.~\ref{fig:eff_case1}(c), we show the total PEI integrated over all photon energies $I_P$ using Eq.~(\ref{eq:total_int}) for the ideal case ($\Gamma_R=\Gamma_{QP}=0$).
Whereas for $L$-photons a sharp peak appears, the emission of $R$-photons occurs over a wider range of $\varepsilon_e$. This is due to the larger hybridization with the $\downarrow$ electron, since $\gamma_1$ is nearly polarized in $\downarrow$-direction (antiparallel to the magnetic field). This hybridization effect is also reflected in Fig.~\ref{fig:eff_case1}(e), where we show the frequency-dependent PEI using Eq.~\eqref{eq:int}. Here, we plot $i_p$ over the QD and photon energy and the gray lines represent the excitation spectrum of the coupled QD-MBSs-system corresponding to the energy differences $\Delta E_e$. The MBSs (horizontal gray lines) are well-separated and have no overlap. The QD states (diagonal gray lines) form anticrossings at resonance due to the finite tunnel coupling $t_1$ which is larger for $\downarrow$ electrons leading to a broader emission of $R$-photons.
The resonances in the PEI for $L$- and $R$-photons correspond to transitions which are illustrated in Fig.~\ref{fig:eff_case1}(a) and \ref{fig:eff_case1}(b), respectively.
The occupied QD levels have to be in resonance with the MBSs so that electrons can be coherently refilled on the QD from the TSC after photon emission. Therefore, only eigenstates which are superpositions of empty and occupied QD levels can contribute to photon emission. When the $\uparrow$ electron is in resonance with the MBSs ($\varepsilon_e=-\Delta_{Z,e}$), the states $\ket{E_1}\approx\ket{\downarrow,1}$ and $\ket{O_1}\approx\ket{\downarrow,0}$ cannot contribute to emission (off-resonant states). However, from the spinless model, see Eq.~\eqref{eq:eff_eigenstates}, we find $\ket{E_{2,3}}\approx\ket{E_{\uparrow\mp}}=(\ket{0,0}\pm\ket{\uparrow,1})/\sqrt{2}$ and $\ket{O_{2,3}}\approx\ket{O_{\uparrow\mp}}=(\ket{0,1}\pm\ket{\uparrow,0})/\sqrt{2}$, which are equal superpositions of an empty and occupied QD state. Therefore, there exists a closed emission cycle for emission of $L$-photons with rate $W^L$, $\ket{E_2}\xleftrightarrow{W^L}\ket{O_2}\xleftrightarrow{W^L}\ket{E_3}\xleftrightarrow{W^L}\ket{O_3}\xleftrightarrow{W^L}\ket{E_2}$, as shown in Fig.~\ref{fig:eff_case1}(a). Since these states are equal superpositions of an empty and occupied QD level, the transitions have the same rates in both directions. Hence, they contribute equally to the stationary state, which leads to maximal PEI.
\begin{figure*}[t]
	\begin{center}
		\includegraphics[width=\textwidth]{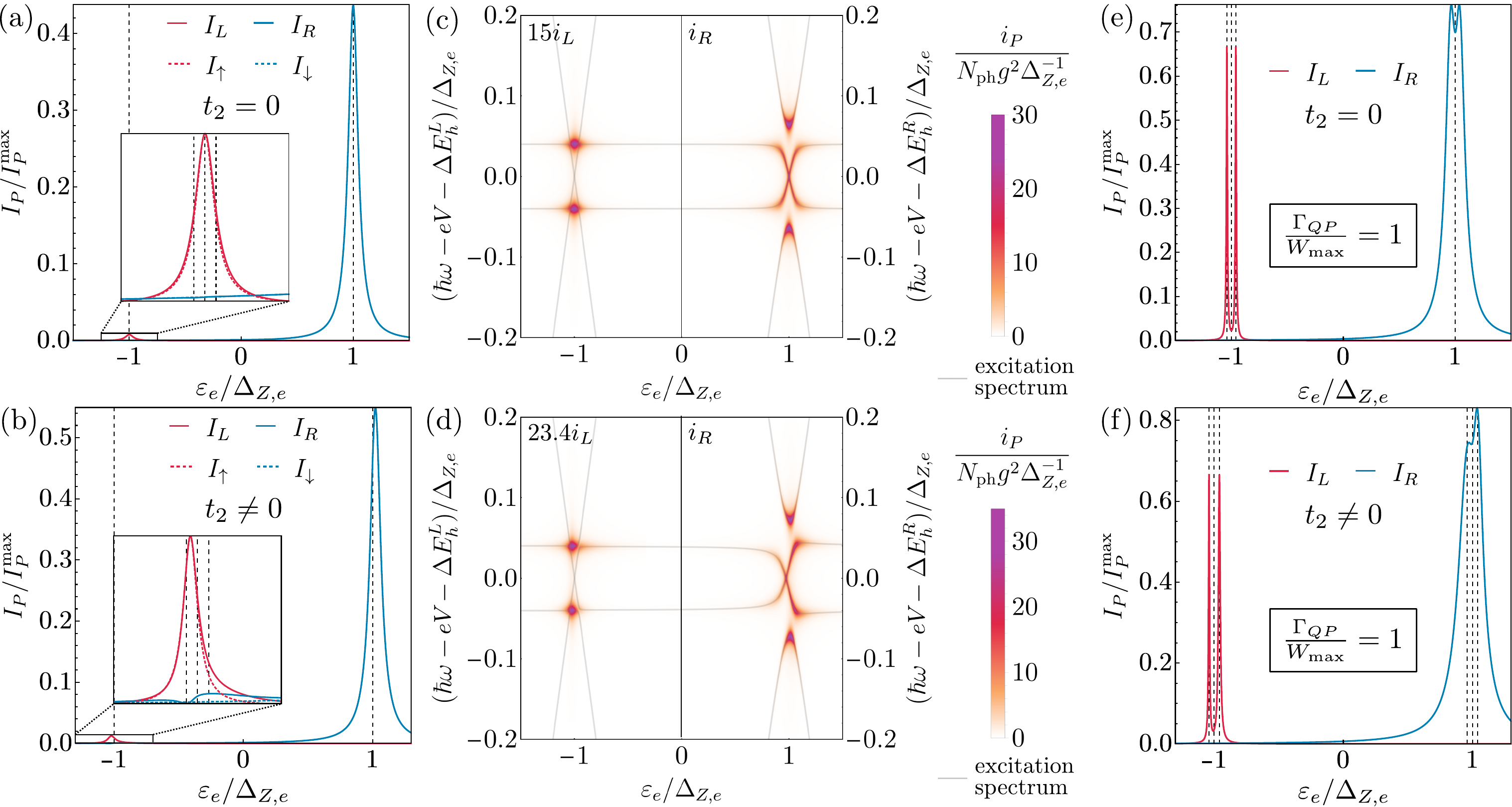}
		\caption{Photon emission for overlapping MBSs. In (a) and (b) we show the total PEI $I_P$ and the PEI $I_\sigma$ calculated from the spinless model for $t_2=0$ and for including a nonlocal coupling $t_2\neq 0$, respectively. In (c) and (d) we show the frequency-dependent PEI $i_P$ over the QD energy $\varepsilon_e$ and the photon energy $\hbar\omega$ for the same parameter regimes as in (a) and (b), respectively , where $\Delta E_h^{L/R}=U_h+\varepsilon_{h\Downarrow/\Uparrow}$. Shown are only transitions from the doubly occupied hole state. Gray lines correspond to the excitation spectrum of the coupled QD-MBSs-system.  In (e) and (f) we show the total PEI $I_P$ where we add a QP with $\Gamma_{QP}=W_{\rm max}$ again for $t_2=0$ and $t_2\neq 0$, respectively. In the case for $t_2=0$, we used $\xi=0.04\Delta_{Z,e}$, $10t_{1\uparrow}=t_{1\downarrow}=0.025\Delta_{Z,e}$. In the case for $t_2\neq0$, we used $\xi=0.04\Delta_{Z,e}$, $10t_{1\uparrow}=t_{1\downarrow}=0.03\Delta_{Z,e}$, $it_{2\uparrow}=0.0009\Delta_{Z,e}$ and $it_{2\downarrow}=-0.009\Delta_{Z,e}$. Other parameters are $\Gamma_{h}= 100W_{\text{max}}$, $\Gamma_{R}=0$.} 
		\label{fig:eff_case2} 
	\end{center}
\end{figure*}
When $\downarrow$ electrons are in resonance ($\varepsilon_e=+\Delta_{Z,e}$), the eigenstates are approximately given by $\ket{E_{1,2}}\approx\ket{E_{\downarrow\mp}}=(\ket{0,0}\pm\ket{\downarrow,1})/\sqrt{2}$ and $\ket{O_{1,2}}\approx\ket{O_{\downarrow\mp}}=(\ket{0,1}\pm\ket{\downarrow,0})/\sqrt{2}$, see Eq.~\eqref{eq:eff_eigenstates}, while $\ket{E_3}\approx\ket{\uparrow,1}$ and $\ket{O_3}\approx\ket{\uparrow,0}$ are now off-resonant. Hence, we find a closed emission cycle $\ket{E_1}\xleftrightarrow{W^R}\ket{O_1}\xleftrightarrow{W^R}\ket{E_2}\xleftrightarrow{W^R}\ket{O_2}\xleftrightarrow{W^R}\ket{E_1}$, illustrated in Fig.~\ref{fig:eff_case1}(b). The two closed emission cycles lead to strong emission of $L$- and $R$-photons at $\varepsilon_e=-\Delta_{Z,e}$ and $\varepsilon_e=+\Delta_{Z,e}$, respectively. Here, $L$- and $R$-PEIs have the same peak height $I_P^{\rm max}$, see Fig.~\ref{fig:eff_case1}(c), which is consistent with the PEI
\begin{align}\label{eq:int_separated}
\frac{I_{\uparrow,\downarrow}}{I_P^{\rm max}}=\frac{2t_{1\uparrow,\downarrow}^2}{2 t_{1\uparrow,\downarrow}^2+(\varepsilon_e\pm\Delta_{Z,e})^2},
\end{align}
in the effectively spinless model, see Eq.~\eqref{eq:eff_int}.
The width of each resonance $2d=2\sqrt{2}|t_{1\uparrow,\downarrow}|$ is proportional to the absolute value of the tunneling amplitude between QD electrons and the MBS $\gamma_1$. Therefore, the relative widths of the PEI peaks give direct access to the MBS spinor-polarization of $\gamma_1$ (for the specific example of a Majorana nanowire, see Sec. \ref{sec:nanowire}).~

If we add spin relaxation with rate $\Gamma_R$ on the QD, we find additional $R$-photons at the resonance with $\uparrow$ electrons ($\varepsilon_e=-\Delta_{Z,e}$), while $I_L$ is decreased. For $\Gamma_R/W_\text{max}=1$, $L$- and $R$-photon emission at $\varepsilon_e=-\Delta_{Z,e}$ is balanced, since half of $\uparrow$ electrons relaxes to $\downarrow$ electrons before being emitted as $L$-photons. But for $\Gamma_R/W_\text{max}>1$, $R$-photon emission is favored. Here, the maximum of $I_R$ is even larger than the maximal PEI without spin relaxation, see Fig.~\ref{fig:eff_case1}(d). At the resonance with $\downarrow$ electrons ($\varepsilon_e=+\Delta_{Z,e}$) $I_R$ stays unchanged. In Fig.~\ref{fig:eff_case1}(a), we show the emission cycle at $\varepsilon_e=-\Delta_{Z,e}$. Here, finite spin relaxation leads to transitions $\ket{E_3(O_3)}\xrightarrow{\Gamma_R}\ket{E_1(O_1)}$ and $\ket{E_2(O_2)}\xrightarrow{\Gamma_R}\ket{E_1(O_1)}$ that preserve the parity. Therefore, the contributions of $\ket{E_1}$ and $\ket{O_1}$ in the stationary state increase, while contributions from the other states decrease. The additional transition paths due to spin relaxation give a new closed emission cycle $\ket{E_1}\xrightarrow{W^R}\ket{O_2}\xrightarrow{\Gamma_R}\ket{O_1}\xrightarrow{W^R}\ket{E_2}\xrightarrow{\Gamma_R}\ket{E_1}$.
Now that the off-resonant states $\ket{E_1}\approx\ket{\downarrow,1}$ and $\ket{O_1}\approx\ket{\downarrow,0}$ have a finite occupation in the steady state, the only possible process for the $\downarrow$ electrons on the QD which do not couple to the MBSs is to recombine to photons. Therefore, spin relaxation can lead to a larger PEI (up to $2I_P^{\rm max}$ for $\Gamma_R\to\infty$) compared to the ideal case where occupied QD states only have 50\% occupation probability in the eigenstates.
However, at the resonance with $\downarrow$ electrons ($\varepsilon_e=+\Delta_{Z,e}$) the PEI remains unchanged compared to the ideal case, see Fig.~\ref{fig:eff_case1}(c). Spin relaxation does not affect the emission cycle between $\ket{E_1},\ket{O_1},\ket{E_2} \text{ and } \ket{O_2}$ at $\varepsilon_e=+\Delta_{Z,e}$, see Fig.~\ref{fig:eff_case1}(b), since $\downarrow$ electrons cannot relax.
We conclude that in the separated MBSs regime, without spin relaxation $\max(I_L)=\max(I_R)$ and the width of the emission peaks is proportional to the tunneling amplitudes between MBSs and QD. A suppressed $L$-photon emission in favor of a strong $R$-photon emission is an indication of a finite $\Gamma_R$ of the order of the optical recombination rate. Usually, spin relaxation times are large compared to photon emission times. Experimentally, spin relaxation rates in QDs are found to be of the order of $\Gamma_R=10^{-5}$ ns$^{-1}$ \cite{Elzerman,HansonReview}.

\subsubsection{Overlapping MBSs}
\begin{figure}[t]
	\begin{center}
		\includegraphics[width=\columnwidth]{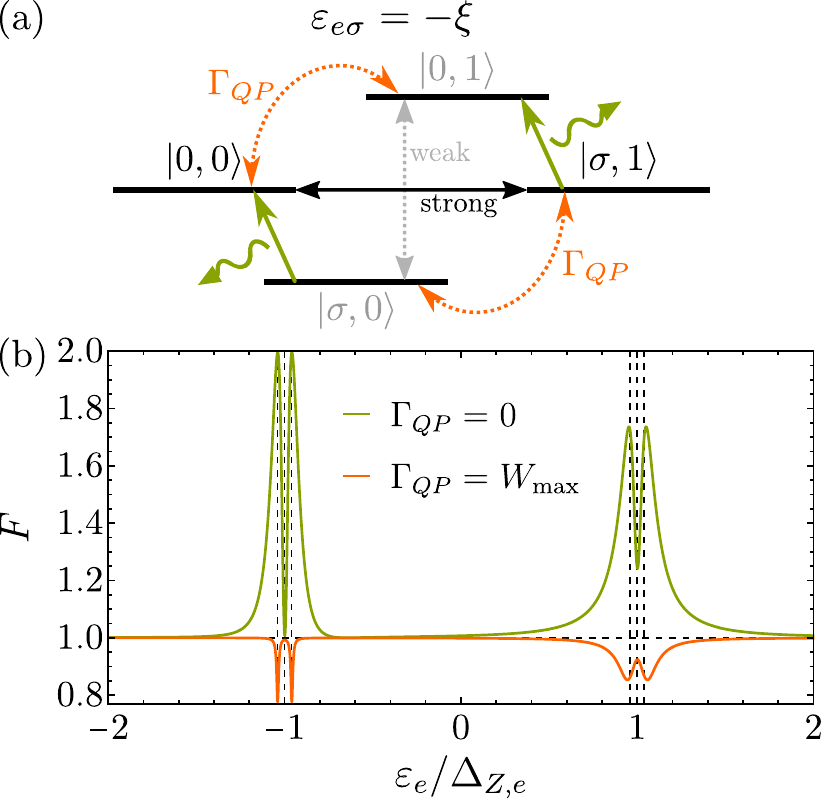}
		\caption{(a) Emission cycles involving the uncoupled QD-MBSs-states $\ket{n_{d\sigma},n_c}$ for $\varepsilon_{e\sigma}=-\xi$. Here, even states are maximally hybridized due to resonant tunneling, whereas odd states are not well hybridized, so PEI is weak and the system stays mostly in the unoccupied QD state $\ket{0,1}$. A finite QP leads to a large PEI, since the weakly hybridized odd states are now connected to the even states via $\Gamma_{QP}$. (b) Fano factor $F$ for overlapping MBSs. We show $F$ over the QD energy $\varepsilon_e$. Without QP the system is in the super-Poissonian regime with $F\geq1$ (green). A finite QP leads to sub-Poissonian noise with $F\leq1$ (orange). These effects are strongest at $\varepsilon_{e\sigma}=\pm\xi$ where odd, respectively, even states hybridize. We used $\xi=0.04\Delta_{Z,e}$, $10t_{1\uparrow}=t_{1\downarrow}=0.025\Delta_{Z,e}$ $t_2=0$, $\Gamma_{h}=100 W_{\rm max}$, $\Gamma_R=0$, $\Gamma_{QP}=0$ (green) and $\Gamma_{QP}=W_{\rm max}$ (orange) (cf. PEIs in Figs.~\ref{fig:eff_case2}(a) and \ref{fig:eff_case2}(e)).} 
		\label{fig:emission_cycle_fano} 
	\end{center}
\end{figure}

Now, we discuss the regime of a finite MBSs splitting $\xi$. First, we consider the case without a tunnel coupling $t_2$ to the second MBS. In this regime, there are two avoided crossings each for even and odd parity in the spectrum at $\varepsilon_e=\pm\Delta_{Z,e}-\xi$ and $\varepsilon_e=\pm\Delta_{Z,e}+\xi$, respectively, see Fig.~\ref{fig:spectrum_regimes}(b).
In Fig.~\ref{fig:eff_case2}(a), we show the total PEI $I_P$ (solid lines) for the full model compared to $I_\sigma$ (dashed lines) for the effectively spinless model, see Eq.~\eqref{eq:eff_int}. Since the spinless model only describes a single spin, whereas the full model includes both spin-states, the difference between the two models comes from a coupling of both spins, which we will discuss in detail in Sec.~\ref{sec:spin_mixing}. For the case discussed here, both models compare very well, see Fig.~\ref{fig:eff_case2}(a), and hence, the spins are to a large degree uncoupled. Here, a finite $\xi$ leads to a reduced maximal $L$- and $R$-PEI while the peak width is increased. Even in the presence of a finite splitting $\xi$, the peaks of $I_L$ and $I_R$ remain located at $\varepsilon_e=-\Delta_{Z,e}$ and $\varepsilon_e=+\Delta_{Z,e}$, respectively.
Furthermore, the maximum of $I_L$ is smaller by one order of magnitude compared to $I_R$, since $|t_{1\uparrow}|<|t_{1\downarrow}|$. The properties of the peaks are well explained by the spinless model, see Eq.~\eqref{eq:Lorentzian_parameters}, where the height and width of the PEI peaks are given by $I_0=2t_{1\sigma}^2/(2t_{1\sigma}^2+\xi^2)$ and $2d=2\sqrt{2t_{1\sigma}^2+\xi^2}$, respectively, and are located at $\varepsilon_{e\sigma}=0$. The decreased peak height and increased peak width compared to the case of $\xi=0$ can be understood best by considering transitions between the product states $\ket{n_d,n_c}$ for the spinless model. At the anticrossing of even states ($\varepsilon_{e\sigma}=-\xi$) without QP, see Fig.~\ref{fig:emission_cycle_fano}(a),
a possible emission cycle is $\ket{\sigma,1}\xrightarrow{W^P}\ket{0,1}\leftrightarrow\ket{\sigma,0}\xrightarrow{W^P}\ket{0,0}\Leftrightarrow\ket{\sigma,1}$. Here, the arrows with $W^P$ indicate optical transitions with emission of a photon with polarization $P=L,R$ and the other arrows correspond to the hybridization between the product states due to tunneling, which can be strong ($\Leftrightarrow$) or weak ($\leftrightarrow$). The strength of the hybridization in this case can be seen from the spectrum in Fig.~\ref{fig:spectrum_regimes}(b) and can be calculated from the occupation probabilities in the eigenstates. If an eigenstate is an equal superposition of an occupied and empty QD level, the occupation probability for each product state is 1/2 which determines the maximal hybridization ($\Leftrightarrow$), otherwise the hybridization is weaker ($\leftrightarrow$). At $\varepsilon_{e\sigma}=-\xi$, the even eigenstates $\ket{E_{\sigma\pm}}$, see Eq.~\eqref{eq:eff_eigenstates_even}, are given by an equal superposition of an empty and occupied QD level, hence the occupation probability $|\langle E_{\sigma\pm}|0,0\rangle|^2=|\langle E_{\sigma\pm}|\sigma,1\rangle|^2=1/2$. However, the odd states $\ket{O_{\sigma\pm}}$ are given by Eq.~\eqref{eq:eff_eigenstates_odd} leading to an occupation probability of $|\langle O_{\sigma\pm}|\sigma,0\rangle|^2=|\langle O_{\sigma\mp}|0,1\rangle|^2=(1\mp\xi/\sqrt{t_{T\sigma}^2+\xi^2})/2$ representing a weaker hybridization. So, for $\xi>0$, the odd states are approximately $\ket{O_{\sigma+}}\approx\ket{0,1}$ and $\ket{O_{\sigma-}}\approx\ket{\sigma,0}$. Because of the weak hybridization between odd states, the system stays mostly in the unoccupied QD state $\ket{0,1}$, see Fig.~\ref{fig:emission_cycle_fano}(a).
At the anticrossing of odd eigenstates ($\varepsilon_{e\sigma}=+\xi$), the system is in the cycle $\ket{\sigma,0}\xrightarrow{W^P}\ket{0,0}\leftrightarrow\ket{\sigma,1}\xrightarrow{W^P}\ket{0,1}\Leftrightarrow\ket{\sigma,0}$, where the stationary state has a large contribution of the unoccupied state $\ket{0,0}$. In both cases, the system stays mostly in a state with no electron on the QD, so the emission is small.
Here, the correlations in the electronic subsystem lead to a super-Poissonian Fano factor $F\geq1$ \cite{Fano1947} of the PEI, see Fig.~\ref{fig:emission_cycle_fano}(b) (green). This effect is due to two different timescales for the emission of photons. Since the occupation probabilities of electrons on the QD in the even and odd parity sector are different, also the recombination rates become different. At the anticrossing of even states ($\varepsilon_e=-\xi$), the rates obey $W^P_{\ket{O_{\sigma+}}\leftarrow \ket{E_{\sigma\pm}}}> W^P_{\ket{O_{\sigma-}}\leftarrow \ket{E_{\sigma\pm}}}$, which leads to two different timescales for emitting a photon from an even state (this also holds for the emission from an odd state where $W^P_{\ket{E_{\sigma\pm}}\leftarrow \ket{O_{\sigma-}}}> W^P_{\ket{E_{\sigma\pm}}\leftarrow \ket{O_{\sigma+}}}$). This effect is strongest at $\varepsilon_{e\sigma}=\pm\xi$, see Fig.~\ref{fig:emission_cycle_fano}(b) (green), where the hybridization between even and between odd states is most different, leading to maximally different recombination rates and a maximum of the Fano factor. If we compare the Fano factor for each spin, it is larger at the resonance for $\uparrow$ electrons ($\varepsilon_e=-\Delta_{Z,e}$) than for $\downarrow$ electrons ($\varepsilon_e=+\Delta_{Z,e}$). Since the coupling $|t_{1\uparrow}|<|t_{1\downarrow}|$ and the hybridization between MBSs and $\uparrow$ electrons is weaker, the system needs more time to return into a state that can emit a photon. Additionally, we can show that there exists a process that leads to bunching of emitted photons on timescales larger than the time $\tau\sim1/\Gamma_{h}$ that the system needs to refill a hole on the QD. This process is reflected in the $g^{(2)}(\tau)$ correlation function \cite{Emary_2012} which measures the correlation between two photon emissions in a given time $\tau$. The details of the calculation of the Fano factor $F$ and $g^{(2)}(\tau)$ are shown in App.~\ref{app:fano}, where we employ a generalized master equation approach \cite{Flindt_2004,Kaiser2007,Dominguez2010}.

Note that for slow holes ($\Gamma_{h}<W_{\rm max}$) the dynamics of the system would be governed by the dynamics in the hole subsystem, since the slowest process determines the dynamics. Here, the Fano factor would be sub-Poissonian, see Fig. \ref{fig:fano_slow_holes}. This can be easily seen from Fig. \ref{fig:transitions}(b). If for example the $\downarrow$ electron is in resonance ($W^L$ is almost 0), mostly $\Uparrow$ holes will recombine to photons. Thus the hole subsystem is in the emission cycle $\ket{\Uparrow\Downarrow}\xrightarrow{W^R}\ket{\Downarrow}\xrightarrow{\Gamma_{h\Uparrow}}\ket{\Uparrow\Downarrow}$, which can be effectively described as a single resonant level between two reservoirs, where the Fano factor is sub-Poissonian and emitted photons are antibunched. An exception constitutes the case where both spins can contribute to photon emission (finite $\Gamma_R$ or in the spin-mixing regime). Then, even for slow holes, the Fano factor can reach the super-Poissonian regime. Additionally, for slow holes, the PEI would be decreased since the maximal PEI is given by $\min(W_{\rm max}/2,\Gamma_{h})$ without spin relaxation.

Since the system changes parity after emitting a photon, the largest PEI is at $\varepsilon_{e\sigma}=0$, see Fig.~\ref{fig:eff_case2}(a), where the emission cycle is $\ket{\sigma,1}\xrightarrow{W^P}\ket{0,1}\xleftrightarrow{*}\ket{\sigma,0}\xrightarrow{W^P}\ket{0,0}\xleftrightarrow{*}\ket{\sigma,1}$ with an intermediate hybridization strength marked by a star ($\xleftrightarrow{*}$). Here, the hybridization between even and between odd states is equally strong leading to the most favorable emission cycle and the maximal PEI in the regime of overlapping MBSs. Here, the maximum is smaller than for the regime of separated MBSs ($\xi=0$) since the eigenstates are not equal superpositions of an empty and an occupied QD level and, as a consequence, the states $\ket{0,1}$ and $\ket{0,0}$ have a larger contribution to the stationary state than $\ket{\sigma,0}$ and $\ket{\sigma,1}$.

Now, we consider a finite coupling $t_2$ to the second MBS in addition to a finite splitting $\xi$. For this regime, the spectrum is shown in Fig.~\ref{fig:spectrum_regimes}(d). Again we show $I_P$ (solid lines) compared to $I_\sigma$ (dashed lines) from the spinless model, which shows only small deviations from $I_P$, see Fig.~\ref{fig:eff_case2}(b), and will not be discussed here. Compared to the case of $t_2=0$, the PEI peak for $I_L$ is shifted in negative $\varepsilon_e$ direction, whereas the peak of $I_R$ is shifted to larger $\varepsilon_e$. The shift of the emission peaks only occurs for $\xi\neq0$ and $t_2\neq0$ ($t_{P\sigma}\neq t_{T\sigma}$). In particular, the PEI peaks are shifted in negative (positive) $\varepsilon_e$ direction if $|t_{P\sigma}|<|t_{T\sigma}|$ ($|t_{P\sigma}|>|t_{T\sigma}|$) where the maximal shift is $\xi$, since the location of the peaks is given by Eq.~\eqref{eq:int_shift} which evaluates to $\varepsilon_0=-2 t_{1\sigma}(it_{2\sigma})\xi/(t_{1\sigma}^2+(it_{2\sigma})^2)$. For the choice of $t_{1\uparrow},it_{2\uparrow},\xi>0$ in Fig.~\ref{fig:eff_case2}(b), we have $|t_{P\uparrow}|<|t_{T\uparrow}|$, such that the $I_L$-peak is shifted in negative $\varepsilon_e$ direction. The shift can be understood from the fact that now the coupling between even parity states is different from the coupling between odd parity states. Here, the hybridization between odd states is stronger than between even states at $\varepsilon_e=-\Delta_{Z,e}$ and since the largest PEI is given where even and odd states can emit photons equally likely, the peak is shifted towards the anticrossing of the more weakly hybridized even states.
At the resonance with $\downarrow$ electrons ($\varepsilon_e=+\Delta_{Z,e}$), the peak is shifted to larger $\varepsilon_e$, since for $t_{1\downarrow},\xi>0$ and $it_{2\downarrow}<0$, we have $|t_{P\downarrow}|>|t_{T\downarrow}|$. Here, the hybridization between even states is stronger than between odd states at resonance, so that the peak is shifted towards the anticrossing of odd states. The emission peak shift is a signature of the nonlocality of the MBSs as it only occurs for finite $t_2$. The effect of a finite $t_2$ can also be seen in the frequency-dependent PEI. In Figs. \ref{fig:eff_case2}(c) and \ref{fig:eff_case2}(d), we show $i_P$ for overlapping MBSs, where the MBSs have a finite splitting (difference between horizontal lines) and $i_L \ll i_R$ (note the different scales). Whereas without $t_2$ the excitation spectrum resembles a bowtie form \cite{Prada2017}, where photon emission is symmetric around the resonances, see Fig. \ref{fig:eff_case2}(c), a finite $t_2$ leads to an asymmetric excitation spectrum and shifts the emission maximum to the less hybridized states (smaller anticrossing), see Fig. \ref{fig:eff_case2}(d).
Note that in the absence of QP the PEI for $t_{1\sigma}=\pm i t_{2\sigma}$ would be zero, since this case would correspond to the coupling to an ordinary complex fermion where superconductivity is effectively suppressed \cite{Schuray2017}.

We now investigate the influence of QP. In Fig.~\ref{fig:eff_case2}(e), we show the total PEI for $\xi\neq0$, $t_2=0$ and $\Gamma_{QP}=W_{\rm max}$. For finite $\Gamma_{QP}$, the total PEI calculated with the spinless model without nonlocal coupling $t_2$ ($t_{P\sigma}=t_{T\sigma}=t_{1\sigma}$) is given by
\begin{align}\label{eq:int_QP}
\frac{I_{\sigma}}{I_P^{\rm max}}=\frac{D_1+D_2\varepsilon_{e\sigma}^2}{D_3+(\varepsilon_{e\sigma}^2-D_4^2)^2}.
\end{align}
The coefficients $D_i,$ $i=1,...,4$ are shown in App.~\ref{app:int_QP}. Note that $I_\sigma$ can never exceed the maximal PEI $I_P^{\rm max}$. At the resonances ($\varepsilon_{e\sigma}=0$), the total PEI simplifies to
\begin{align}\label{eq:int_QP_0}
\frac{I_{\sigma}(\varepsilon_{e\sigma}=0)}{I_P^{\rm max}}=\frac{2t_{1\sigma}^2(1+2\tilde{\Gamma}_{QP})}{2t_{1\sigma}^2(1+2\tilde{\Gamma}_{QP})+\xi^2},
\end{align}
with the dimensionless rate $\tilde{\Gamma}_{QP}=\Gamma_{QP}/W_{\rm max}$. Compared to the case without QP, it effectively changes the tunneling amplitudes $\tilde t_{1\sigma}=t_{1\sigma}\sqrt{1+2\tilde\Gamma_{QP}}$ at $\varepsilon_{e\sigma}=0$ and the PEI $I_\sigma(\varepsilon_{e\sigma}=0)$ increases linearly with $\tilde\Gamma_{QP}$, see Eq.~\eqref{eq:int_QP_0}. 
Furthermore, under the condition that 
\begin{align}\label{eq:QP-condition}
\tilde{\Gamma}_{QP}>\frac{t_{1\sigma}^2}{3\xi^2-2t_{1\sigma}^2}> 0,
\end{align}
the function $I_{\sigma}$ has two peaks which are symmetric around $\varepsilon_{e\sigma}=0$ and split by
\begin{align}\label{eq:split_peaks}
\Delta E_{\rm split}=\left|2\xi-\frac{(1+2\tilde{\Gamma}_{QP})^2}{4 \tilde{\Gamma}_{QP}^2\xi^3}t_{1\sigma}^4+\mathcal{O}(t_{1\sigma}^6)\right|.
\end{align}
Otherwise, $I_{\sigma}$ has one peak at $\varepsilon_{e\sigma}=0$. In Fig.~\ref{fig:eff_case2}(e), $I_L$ has two peaks and $I_R$ has two broader overlapping peaks. The peaks for a given photon polarization are approximately split by $2|\xi|$, see Eq.~\eqref{eq:split_peaks}, and are symmetric around the resonance $\varepsilon_e=\pm\Delta_{Z,e}$, as $t_2=0$ in this case. Also the maximal PEI is strongly enhanced, especially at $\varepsilon_e=-\Delta_{Z,e}\pm\xi$, if we compare this to the case without QP, see Figs.~\ref{fig:eff_case2}(a) and \ref{fig:eff_case2}(e). This difference to the ideal case can be understood with the emission cycle at $\varepsilon_{e\sigma}=-\xi$, illustrated in Fig.~\ref{fig:emission_cycle_fano}(a). Quasiparticle poisoning leads to additional paths, so that the new emission cycle $\ket{\sigma,1}\xrightarrow{W^P}\ket{0,1}\xleftrightarrow{\Gamma_{QP}}\ket{0,0}\Leftrightarrow\ket{\sigma,1}$ is created. Now, the weak hybridization between odd states at $\varepsilon_{e\sigma}=-\xi$ can be overcome by changing the parity via QP, which strongly enhances the PEI. Likewise, at $\varepsilon_{e\sigma}=+\xi$ where even states are weakly hybridized, the new emission cycle $\ket{\sigma,0}\xrightarrow{W^P}\ket{0,0}\xleftrightarrow{\Gamma_{QP}}\ket{0,1}\Leftrightarrow\ket{\sigma,0}$ enhances the emission. Here, the Fano factor becomes sub-Possonian, see Fig.~\ref{fig:emission_cycle_fano}(b) (orange). Again this effect is strongest at QD energies where even or odd states are strongly hybridized. Since QP changes the occupation of the nonlocal fermion, see Fig. \ref{fig:emission_cycle_fano}(a), even and odd parity eigenstates are mixed. Since parity is not conserved, the electronic subsystem becomes effectively a two-level system with $\ket{0}$ and $\ket{\sigma}$ on the QD, so that the emission cycle is effectively $\ket{0}\Leftrightarrow\ket{\sigma}\xrightarrow{W^P}\ket{0}$. For such a single resonant level system, the Fano factor is sub-Poissonian and emitted photons are antibunched for all times $\tau$. This effect emerges if $\tilde{\Gamma}_{QP}>|t_{1\sigma}|/2\sqrt{t_{1\sigma}^2+\xi^2}$. Further details are shown in App. \ref{app:fano}.
\begin{figure*}[t!]
	\begin{center}
		\includegraphics[width=.9\textwidth]{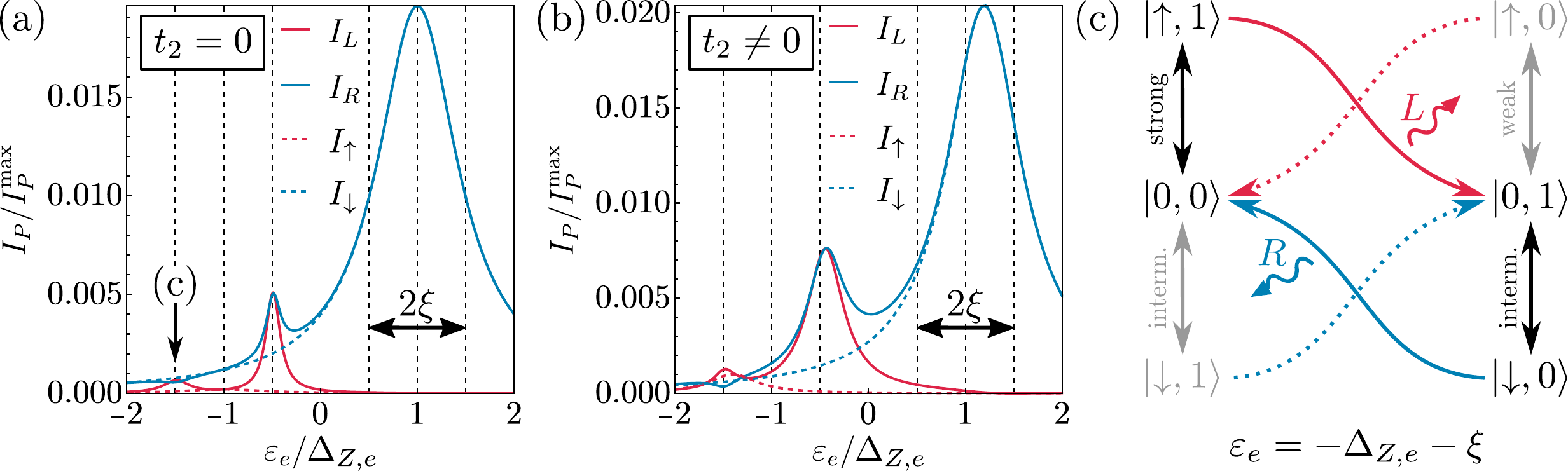}
		\caption{Photon emission in the spin-mixing regime. 
			(a,b) The total PEI $I_P$ shows two additional peaks at $\varepsilon_e=-\Delta_{Z,e}\pm\xi$ due to a spin-mixing effect which do not appear in the spinless model PEI $I_\sigma$. We used $\xi=0.5 \Delta_{Z,e}$, $10t_{1\uparrow}=t_{1\downarrow}=0.05\Delta_{Z,e}$, and respectively $t_2=0$ in (a) and $i t_{2\uparrow}=0.01\Delta_{Z,e}$, $i t_{2\downarrow}=-0.01\Delta_{Z,e}$ in (b). Further parameters are $\Gamma_{h}= 100W_{\text{max}}$ and $\Gamma_{QP}=\Gamma_{R}=0$. (c) In the emission cycle for $\varepsilon_e=-\Delta_{Z,e}-\xi$, we show the strength of hybridization (black/gray arrows) between the states $\ket{n_{d\sigma},n_c}$ and optical transitions with emission of $L$- (red arrows) or $R$-photons (blue arrows). In the spin-mixing regime, the most likely emission cycle is denoted by red/blue solid arrows, whereas the dotted paths are rather unlikely.} 
		\label{fig:eff_case3} 
	\end{center}
\end{figure*}

In Fig.~\ref{fig:eff_case2}(f), we show the PEI for finite $\Gamma_{QP}$ and $t_2$. For $I_L$, $t_2$ changes the relative width of the two peaks at $\varepsilon_e=-\Delta_{Z,e}\pm\xi$. Here, odd states are more hybridized at $\varepsilon_e=-\Delta_{Z,e}+\xi$ than even states at $\varepsilon_e=-\Delta_{Z,e}-\xi$. QP enhances the emission more where states of one parity hybridize less than states of the other parity because it allows the parity to change. Since at $\varepsilon_e=-\Delta_{Z,e}+\xi$ even states hybridize less than odd states, the largest effect of QP can be seen around this resonance.  Here, the $I_L$-peak at $\varepsilon_e=-\Delta_{Z,e}+\xi$ is more broadened, see Fig. \ref{fig:eff_case2}(f). For $I_R$, the peak height of the two peaks at $\varepsilon_e=+\Delta_{Z,e}\pm\xi$ is different now and the peaks are overall broadened. Due to the coupling to the second MBS, the PEI gets shifted to larger $\varepsilon_e$ which merges together with the QP into a double-peak PEI, where the peak at larger $\varepsilon_e$ is higher. The shift and asymmetry of the  double-peak PEI is much less visible for $I_L$, since there the PEI is much weaker without QP, so that QP is the leading mechanism.

In summary, for overlapping MBSs the PEI peaks are decreased in height and broadened compared to the case of separated MBSs, since even and odd states are separated in energy by $\xi$. The coherent coupling between electrons on the QD and the MBSs leads to a super-Poissonian Fano factor due to two different timescales for recombination rates in the even and odd subspace and for timescales larger than $\tau\sim1/\Gamma_{h}$ emitted photons are bunched.
A finite $t_2$ leads to a shift of the emission peaks which is a signature of nonlocality of the MBSs. Additionally, QP leads to a significant increase of the PEI and can lead to the emergence of two peaks that are split by $\sim2|\xi|$, as a weak hybridization between states can be overcome by changing the parity of the system via QP. A similar splitting and enhancement of a transport peak due to QP has been also reported in charge transport through a QD coupled to a Majorana nanowire \cite{Leijnse2011}. Furthermore, QP leads to a sub-Poissonian Fano factor and antibunching of photons for all times $\tau$. We conclude that in this regime, a super-Poissonian Fano factor is a signature of a Majorana system where $\tilde{\Gamma}_{QP}<|t_{1\sigma}|/2\sqrt{t_{1\sigma}^2+\xi^2}$ and thus the parity in the system is conserved on timescales of the order of the photon emission. The timescale for QP becomes comparable to the photon emission time for a light-matter-interaction energy of $g\sim\sqrt{\hbar\Gamma_{QP}/N_{\rm ph}}$. A typical QP rate stated in the literature is~ $\Gamma_{QP}=10^{-3}$~ns$^{-1}$ \cite{Rainis}.

\subsection{Spin-mixing regime}\label{sec:spin_mixing}
When $\Delta_{Z,e}\gtrsim t_1$ and/or $\Delta_{Z,e}\gtrsim \xi$, the resonances of $\uparrow$- and $\downarrow$ electrons in the total PEI have a significant overlap. In this regime, the model of decoupled spins in Eq.~\eqref{eq:spinless} is not expected to be a good approximation, since the two spins are strongly coupled via the MBSs. In this subsection, we consider the example of a MBS splitting $\xi$ that is smaller but on the order of $\Delta_{Z,e}$. Note that spin-mixing effects also occur for smaller splittings, if the tunnel couplings are large enough. In Fig.~\ref{fig:eff_case3}(a), we show the total PEI $I_P$ (solid lines) compared to the spinless model $I_\sigma$ (dashed lines) for the ideal case with a finite splitting $\xi$ and $t_2=0$. Here, $I_R$ has a broad peak centered around $\varepsilon_e=+\Delta_{Z,e}$, a small peak at $\varepsilon_e=-\Delta_{Z,e}+\xi$ and a dip at $\varepsilon_e=-\Delta_{Z,e}-\xi$. $I_L$ has two peaks at $\varepsilon_{e}=\Delta_{Z,e}\pm\xi$. Interestingly, $I_L$ and $I_R$ are nearly equal at $\varepsilon_e=-\Delta_{Z,e}\pm\xi$. In the regime of $\Delta_{Z,e}\gg t_1,t_2,\xi$ with $\Gamma_{QP}=0$, the PEI had only one peak for each photon polarization. The extra peaks have to emerge due to the coupling of different spins which leads to an emission cycle where $L$- and $R$-photons both are emitted, as illustrated in Fig.~\ref{fig:eff_case3}(c), where red and blue arrows correspond to emission of $L$- and $R$-photons, respectively.

Without spin-mixing, emission of $L$-photons at $\varepsilon_e=-\Delta_{Z,e}-\xi$ would be weak, since this would correspond to the emission cycle in Fig.~\ref{fig:emission_cycle_fano}(a). In contrast, in the spin-mixing regime, tunneling of $\downarrow$ electrons is now also present. Hence, a new emission cycle is possible, $\ket{\uparrow,1}\xrightarrow{W^L}\ket{0,1}\xleftrightarrow{*}\ket{\downarrow,0}\xrightarrow{W^R}\ket{0,0}\Leftrightarrow\ket{\uparrow,1}$, see solid red and blue lines in Fig.~\ref{fig:eff_case3}(c). Here, one $L$-photon and one $R$-photon are emitted in a full cycle. This leads to equal PEIs $I_L$ and $I_R$. At $\varepsilon_e=-\Delta_{Z,e}+\xi$, spin-mixing leads to the new cycle $\ket{0,0}\xleftrightarrow{*}\ket{\downarrow,1}\xrightarrow{W^R}\ket{0,1}\Leftrightarrow\ket{\uparrow,0}\xrightarrow{W^L}\ket{0,0}$. Note that these cycles are the most likely paths, but other paths (dotted arrows in Fig.~\ref{fig:eff_case3}(c)) are also possible, which can lead to $I_L\neq I_R$ depending on the relation between the couplings $t_{1\sigma}$ and the splitting $\xi$. The peaks of $I_P$ at $\varepsilon_e=-\Delta_{Z,e}-\xi$ are smaller than at $\varepsilon_e=-\Delta_{Z,e}+\xi$, since the further the QD energy is detuned from the resonance at $\varepsilon_e=+\Delta_{Z,e}$, the weaker is the hybridization with $\downarrow$ electrons.

A finite nonlocal coupling $t_2$ can influence the spin-mixing effect, since it changes the tunneling amplitudes between even and between odd states, which affects the emission paths and can lead to different emission peak widths and heights. In Fig.~\ref{fig:eff_case3}(b), we show the total PEI in the spin-mixing regime with a nonlocal coupling. Now, the broad $I_R$ peak is shifted to larger $\varepsilon_e$ and both the height and width of the two spin-mixing peaks are modified compared to the case of $t_2=0$ in Fig.~\ref{fig:eff_case3}(a). Interestingly, at $\varepsilon_e=-\Delta_{Z,e}-\xi$, we observe that $I_L>I_R$. The additional emission of $L$-photons corresponds to the path $\ket{0,1}\leftrightarrow\ket{\uparrow,0}\xrightarrow{W^L}\ket{0,0}$, see dotted red line in Fig.~\ref{fig:eff_case3}(c), which is now more likely. We note that spin relaxation in the case of decoupled spins can lead to $R$-photon emission at the resonance with $\uparrow$ electrons as well. However, in the spin-mixing regime both spin species contribute to photon emission so that $L$- and $R$-photon emission reinforce each other which increases both the $L$- and $R$-PEI.

\section{Majorana Nanowire}\label{sec:nanowire}
In this section, we comment on how our proposed $pn$ junction could be implemented using a semiconducting nanowire with Rashba spin-orbit coupling in the presence of a magnetic field along the wire direction. The wire has three separated regions, see Fig.~\ref{fig:combi_setup}(a). The $n$ side is in proximity to an s-wave SC forming a TSC, where in the topologically non-trivial phase MBSs emerge at the ends, while the $p$ side is a normal conducting reservoir for holes. The QD confines electrons and holes and is embedded in the $pn$ junction. Electrons on the QD are coupled to the MBSs at the left side of the $n$ side TSC and holes can be refilled from the $p$ side. Via electron-hole recombination, photons of $L$ or $R$ polarization are emitted in the wire direction, since the magnetic field is applied along the wire direction. The location of the QD on the nanowire axis exhibits an efficient photon emission in wire direction, if the wire acts as a waveguide for the emitted photons \cite{ReimerNW,Versteegh}.

For a Zeeman energy $\Delta_Z>\sqrt{E_F^2+\Delta_S^2}$ with $E_F$ the Fermi energy of the wire and $\Delta_S$ the induced superconducting (s-wave) pairing the $n$ side is in the topologically non-trivial phase, where MBSs emerge at the ends of the TSC \cite{Lutchyn,Oreg}. Note that $\Delta_Z$ in the wire and $\Delta_{Z,e}$ on the QD can be different, either due to different g-factors (different materials), different external magnetic fields or by implementing the Zeeman splitting in the wire in situ via the exchange interaction with a ferromagnetic insulator \cite{Vaitiekenas2021}. Since we neglected the SC continuum in our considerations, we need $\Delta_{Z,e} < \Delta_S$ in order to have both spin-levels within the SC gap $\Delta_S$ \footnote{If $\Delta_{Z,e} > \Delta_S$ and the $\uparrow$ electron is in resonance with the MBSs, the coherent tunneling of $\uparrow$ electrons can become blocked by the tunneling of a $\downarrow$ electron onto the empty QD by the simultaneous creation of a quasiparticle in the SC. Phenomenologically, this process is similar to the spin relaxation process with rate $\Gamma_R$ with the difference that the parity of the low-energy states does switch.}. This condition can be relaxed if the coupling energy of the $\downarrow$ electron QD level to the continuum states above $\Delta_S$ is small (i.e. smaller than $\hbar W_{\rm max}$ and $\min(|t_{P\uparrow}|,|t_{T\uparrow}|)$) \footnote{This coupling energy can be smaller than $\min(|t_{P\uparrow}|,|t_{T\uparrow}|)$ since the Majorana wave function is localized within the superconducting coherence length which is much smaller than the quantization length of the continuum states.}. One can obtain the electronic components of the Majorana wave functions $u_{\sigma}^{(i)}$ at the left side of the $n$ side wire for $\gamma_i$, $i=1,2$, $\sigma= \ \uparrow,\downarrow$. By choosing the magnetic field in $z$-direction and the Rashba spin-orbit coupling in $y$-direction, the spins of the electronic components of the MBSs are polarized in the $z-x$ plane, so that their spin at the left end can be parameterized by a single angle
\begin{align}
\label{eq:MBS_angle}\Theta_i=2\arccot\left(\frac{u_{\uparrow}^{(i)}}{ u_{\downarrow}^{(i)}}\right),
\end{align}
$i=1,2$ \cite{Sticlet2012,Prada2017,Schuray2018}.
The tunneling amplitudes between QD electrons and MBSs are $t_{i\sigma}\propto u_\sigma^{(i)}$ \cite{Prada2017,Hoffman2017,Schuray2018}, which correspond to the tunneling amplitudes we introduced in Eq. \eqref{eq:HamEl}. Since $\gamma_1$ ($\gamma_2$) is exponentially localized at the left (right) end \cite{Klinovaja2012,DasSarma2012}, the coupling to $\gamma_2$ is only finite if its wave function can reach the left end. The strength of the couplings can be tuned by the strength of the tunneling barrier between QD and wire. 

For sufficiently long $n$ side wires, the Majorana splitting $\xi$ and the coupling $t_2$ to the more distant MBS $\gamma_2$ are negligibly small, which corresponds to the regime of separated MBSs. From the effectively spinless model (decoupled spins), we find from Eq.~\eqref{eq:int_separated} a relation between the width $2d$ of the emission peaks and the Majorana angle $\Theta_1$ in Eq.~\eqref{eq:MBS_angle} as
\begin{align}\label{eq:Majorana_angle}
\Theta_1&=2\arccot\left( \frac{d(I_L)}{d(I_R)}\right).
\end{align}
For shorter wires, the splitting $\xi$ \cite{DasSarma2012,Rainis2013} and the tunneling amplitudes $t_{2\sigma}$ \cite{Prada2017} become finite, which corresponds to the regime of overlapping MBSs. Additionally, $\xi$ and the tunneling amplitudes $t_{i\sigma}$ oscillate with increasing $\Delta_Z$.
By decreasing the wire length further or tuning the tunnel barrier between QD and wire, one can also reach the spin-mixing regime. Here, the Zeeman energy on the QD has to be tuned such that the Majorana splitting and/or tunneling amplitudes are smaller but on the order of the Zeeman energy, i.e. $\Delta_{Z,e}\gtrsim \xi$ and/or $\Delta_{Z,e}\gtrsim t_1$.

\section{Conclusion}\label{sec:conclusion}
We considered an effective model for the coupling of two Majorana bound states (MBSs) to electrons on a quantum dot (QD). We introduced holes which do not interact with the MBSs but are coupled to a normal conducting reservoir to calculate electron-hole recombination with emission of left ($L$) and right ($R$) polarized photons governed by selection rules. We used a master equation to calculate the dynamics of the system, where we included spin relaxation and quasiparticle poisoning (QP). 

The photon emission intensity (PEI) behaves differently depending on the MBSs regimes. For spatially separated MBSs, the height of the PEI obeys $\max(I_L)=\max(I_R)$ and the width is proportional to the tunneling amplitudes between electrons on the QD and MBSs. When the wave functions of the MBSs 
overlap significantly, the PEI peaks obey $\max(I_L)\neq\max(I_R)$ and are shifted away from resonance if a finite tunnel coupling to the more distant MBS is present, which is a signature of the nonlocality of a pair of MBSs. Additionally, we showed that for overlapping MBSs the system dynamics leads to bunching of photon emission events resulting in a super-Poissonian Fano factor ($F\geq1$). The reason can be linked to two different timescales for the emission of photons, since the hybridization between MBSs and QD states in the even and odd parity sector is different and bunching of emitted photons appears on timescales larger than the hole refilling time $\tau\sim1/\Gamma_{h}$. Furthermore, a finite spin relaxation on the QD leads to an enhanced emission of $R$-photons at the tunneling resonance of $\uparrow$ electrons, which can be larger than the maximal emission without spin relaxation. A finite QP leads to an enhancement of the PEIs and can split the $L$- and $R$ photon resonances into two peaks each, separated in energy 
approximately by $|2\xi|$. Due to mixing of the parities via QP, the system is in the sub-Poissonian regime ($F\leq1$) and emitted photons are always antibunched. We conclude that in this regime, a super-Poissonian Fano factor is a signature of a Majorana system where QP is slow enough, so that the parity in the system is conserved on timescales on the order of the photon emission. Moreover, for larger tunnel couplings between the QD and MBSs and/or larger Majorana splittings, the two spin levels on the QD become effectively coupled by the MBSs which leads to spin-mixing effects in the PEI. For a large MBSs splitting and $|t_{1\downarrow}|>|t_{1\uparrow}|$, this gives rise to additional $R$-photons at the tunneling resonance for $\uparrow$ electrons where one has to carefully distinguish between the cases of spin-mixing and spin relaxation, as we can find $R$-photons at the resonance with $\uparrow$ electrons in both cases.

Our proposed $pn$ junction could be implemented using a semiconducting Rashba nanowire proximitized to 
an s-wave superconductor ($n$ side) and subjected to a Zeeman field, that hosts a MBS 
at each end. The QD can be embedded in a $pn$ junction between the $n$ side TSC and a normal conducting $p$ side reservoir. For long wires, the MBSs have only little overlap and are well-separated. In that case, the Majorana spin angle of $\gamma_1$ can be directly calculated from the emission peak widths. For shorter wires, also the regimes of overlapping MBSs and spin-mixing can be reached by tuning the Zeeman energy appropriately.

The mapping of MBS-properties to photons is an alternative route to transport measurements to investigate MBSs. Furthermore, the detection provides direct access to the polarization of the photons and allows to distinguish between QD electrons of different spins, which is not readily achievable in charge transport. With our proposed scheme we can clearly distinguish the different MBSs regimes and get access to parameters like the Majorana spin angle, splitting of the MBSs and the order of QP times in the system. 
Given the existing successful combination of nanowires, SCs and QDs in the Majorana realm~\cite{Deng2016,Deng2018} as well as the integration of optically active QDs into nanowires~\cite{Minot} with spin-selective PEI~\cite{vanWeert2009} and directional emission of photons~\cite{ReimerNW,Versteegh}, we believe that our proposal is experimentally feasible.

Currently, it is discussed how to distinguish MBSs from non-topological zero-energy states. Our proposal has the advantage that we have direct access to the spin structure of the probed states from the resulting polarization resolved PEIs. With our setup we can probe the Majorana overlap as well as their nonlocality, which are crucial for the demonstration of MBSs. But since topologically trivial zero-energy states such as quasi-MBSs \cite{Vuik2019} can have a non-trivial spin structure as well, our local probe cannot unambiguously distinguish these states from MBSs (for a review on this topic, see Ref. \cite{Prada2020}).

During writing of the manuscript we became aware of a related publication \cite{Ricco2022}, where a spinless optical QD connected to MBSs in a cavity is considered. 
The QD is dressed by the coupling to the cavity photons which, in turn, influences the charge transport (without a $pn$ junction) through the QD, all very different from our scheme.

\section*{Acknowledgments}
We thank V. Zwiller for comments on the manuscript and F. Dominguez for fruitful discussions related to the shot noise considerations. LB, DF, and PR acknowledge financial support by the Deutsche Forschungsgemeinschaft (DFG, German Research Foundation) within the framework of Germany’s Excellence Strategy-EXC-2123 QuantumFrontiers-390837967 and LB and PR acknowledge financial support from “Nieders\"achsisches Vorab” through “Quantum- and Nano-Metrology (QUANOMET)” initiative within the project NL-2.
	%
	%
\appendix 
\begin{figure*}[t!]
	\centering
	\includegraphics[width=\textwidth]{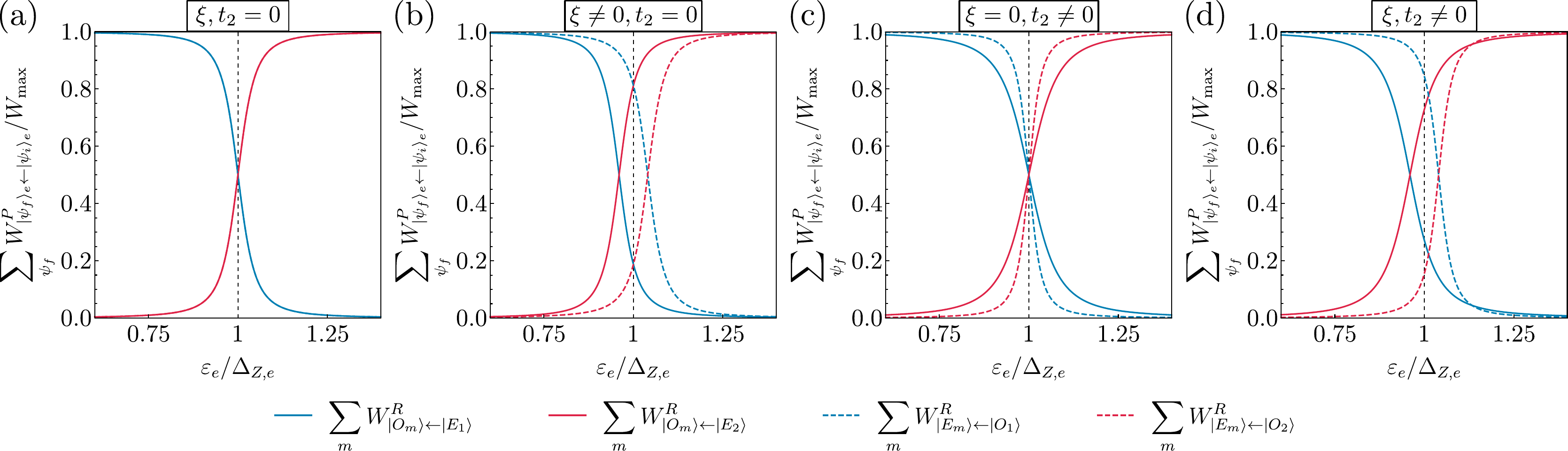}
	\caption{Relevant recombination rates over QD energy $\varepsilon_e$ for different MBSs regimes at the resonance with $\downarrow$ electrons ($\varepsilon_e=\Delta_{Z,e}$). (a) Separated MBSs. Even and odd states are degenerate, which leads to the same rates for each parity. Parameters are $\xi=0$, $10t_{1\uparrow}=t_{1\downarrow}=0.025\Delta_{Z,e}$, $t_2=0$. (b) Overlapping MBSs with finite splitting. Parameters are $\xi=0.04\Delta_{Z,e}$, $10t_{1\uparrow}=t_{1\downarrow}=0.025\Delta_{Z,e}$, $t_2=0$. (c) Overlapping MBSs with finite coupling to $\gamma_2$. Parameters are $\xi=0$, $10t_{1\uparrow}=t_{1\downarrow}=0.03\Delta_{Z,e}$, $it_{2\uparrow}=0.0012\Delta_{Z,e}$, $it_{2\downarrow}=-0.012\Delta_{Z,e}$. (d) Overlapping MBSs with finite splitting and finite coupling $t_2$. Parameters are $\xi=0.04\Delta_{Z,e}$, $10t_{1\uparrow}=t_{1\downarrow}=0.03\Delta_{Z,e}$, $it_{2\uparrow}=0.0009\Delta_{Z,e}$, $it_{2\downarrow}=-0.009\Delta_{Z,e}$.} 
	\label{fig:rates} 
\end{figure*}
\section{Recombination rates}\label{app:rates}
Electrons and holes on the QD recombine to photons via optical recombination. For the total recombination rates in Eq.~\eqref{eq:changerate} we have to calculate all possible matrix elements $M_{f,i}^P$ in Eq.~\eqref{eq:matrixelements}. The eigenstates of the full system are product states of the form $\ket{n_{d\sigma},n_c}\times\ket{n_{h\sigma}}\times\ket{n_{\rm ph}}$, where $\hat{n}_{\rm ph}=a_{k,P}^\dagger a_{k,P}$ is the occupation number operator of photons.
The hole eigenstates are simply given by $\ket{\psi}_h=\{\ket{0}_h,\ket{\Uparrow},\ket{\Downarrow},\ket{\Uparrow\Downarrow}\}$. 
We can show that transitions $\ket{\Uparrow\Downarrow}\xrightarrow{W^L(W^R)}\ket{\Uparrow(\Downarrow)}$ and $\ket{\Downarrow(\Uparrow)}\xrightarrow{W^L(W^R)}\ket{0}_h$ in the hole subsystem lead to the same recombination rates and since $[d_\sigma^{(\dagger)},a_{k,P}^{(\dagger)}]=[h_\sigma^{(\dagger)},a_{k,P}^{(\dagger)}]=0,$ we can simplify
\begin{equation}\label{eq:rates_short}
	\begin{split}
W^L_{\ket{\psi_f}_e \ket{\Uparrow} \leftarrow \ket{\psi_i}_e \ket{\Uparrow\Downarrow}}=W^L_{\ket{\psi_f}_e \ket{0}_h \leftarrow \ket{\psi_i}_e \ket{\Downarrow}}=W^L_{\ket{\psi_f}_e  \leftarrow \ket{\psi_i}_e},\\
W^R_{\ket{\psi_f}_e \ket{\Downarrow} \leftarrow \ket{\psi_i}_e \ket{\Uparrow\Downarrow}}=W^R_{\ket{\psi_f}_e \ket{0}_h \leftarrow \ket{\psi_i}_e \ket{\Uparrow}}=W^R_{\ket{\psi_f}_e  \leftarrow \ket{\psi_i}_e }.
\end{split}
\end{equation}
Therefore, we suppress the hole index in the recombination rates and the photon polarization only appears as an index.
The electronic eigenstates have a fixed parity $\ket{\psi}_e=\ket{E_m(O_m)}$ and are superpositions of the product states $\ket{n_{d\sigma},n_c}$ of QD electrons and the nonlocal fermion. So even and odd eigenstates are respectively given by
\begin{subequations}\label{eq:eigenstate_components}
\begin{align}
\ket{E_m}=\varphi_{1m}^e\ket{0,0}+\varphi_{2m}^e\ket{\uparrow,1}+\varphi_{3m}^e\ket{\downarrow,1}+\varphi_{4m}^e\ket{\uparrow\downarrow,0},\\
\ket{O_m}=\varphi_{1m}^o\ket{0,1}+\varphi_{2m}^o\ket{\uparrow,0}+\varphi_{3m}^o\ket{\downarrow,0}+\varphi_{4m}^o\ket{\uparrow\downarrow,1}.
\end{align}
\end{subequations}
Since the master equation is written in terms of the coherently coupled QD-MBSs-states, we have to calculate the contribution of every eigenstate by its components. For the recombination rates in Eq.~\eqref{eq:changerate} we obtain explicitly 
\begin{subequations}\label{eq:rates_explicitly}
\begin{align}
W^L_{\ket{O_m}\leftarrow \ket{E_n}}=\frac{2\pi}{\hbar}\left|g \left(\varphi_{1m}^{o*}\varphi_{2n}^e+\varphi_{3m}^{o*}\varphi_{4n}^e\right)\right|^2 N_{\rm ph},\\
W^R_{\ket{O_m}\leftarrow \ket{E_n}}=\frac{2\pi}{\hbar}\left|g \left(\varphi_{1m}^{o*}\varphi_{3n}^e-\varphi_{2m}^{o*}\varphi_{4n}^e\right)\right|^2 N_{\rm ph},\\
W^L_{\ket{E_m}\leftarrow \ket{O_n}}=\frac{2\pi}{\hbar}\left|g \left(\varphi_{1m}^{e*}\varphi_{2n}^o+\varphi_{3m}^{e*}\varphi_{4n}^o\right)\right|^2 N_{\rm ph},\\
W^R_{\ket{E_m}\leftarrow \ket{O_n}}=\frac{2\pi}{\hbar}\left|g \left(\varphi_{1m}^{e*}\varphi_{3n}^o-\varphi_{2m}^{e*}\varphi_{4n}^o\right)\right|^2 N_{\rm ph}.
\end{align}
\end{subequations}
In Fig.~\ref{fig:rates}, we show the recombination rates over the QD energy $\varepsilon_e$ for different MBSs regimes at the resonance with $\downarrow$ electrons ($\varepsilon_e=\Delta_{Z,e}$).
Away from resonance, the recombination rates are either 0 or $W_{\text{max}}=2\pi |g|^2 N_{\rm ph}/\hbar$ because here electronic eigenstates correspond to uncoupled states $\ket{n_{d\sigma},n_c}$ (off-resonant states). If electrons on the QD are in resonance with the MBSs, it follows for the rates that $0< W^P_{\ket{\psi_f}_e  \leftarrow \ket{\psi_i}_e}< W_{\rm max}$. In this $\varepsilon_e$ range, the eigenstates that contribute to photon emission are superpositions of the uncoupled QD-MBSs-states. Only for those eigenstates photon emission is possible, since a maximal emission rate would lead to zero occupation in the stationary state and thus the contribution to the intensity defined in Eq.~\eqref{eq:int} would be zero. The same holds for a recombination rate of zero which immediately gives zero intensity.
Note that off-resonant states can contribute to emission, if they are coupled to the resonant states via other processes, which is e.g. the case for a finite spin relaxation.

\section{Master equation}\label{app:master}

The dynamics of the system is described by a Pauli master equation for the 32 joint states $\ket{\psi}=\ket{\psi}_e\times \ket{\psi}_h$. We take into account four different rates: The optical recombination rates are given by Eq.~\eqref{eq:rates_explicitly}. The other rates for hole refilling, QP and spin relaxation can be deduced, respectively, by taking the expectation value in the system state $\ket{\psi}$ of the Lindblad superoperators in Eq.~\eqref{eq:Lindblad_hole}, \eqref{eq:Lindblad_QP} and \eqref{eq:Lindblad_relax},
\begin{align}
\bra{\psi}\mathcal{L}_{h\sigma}\ket{\psi}&=\sum_{\psi'} \left(\Gamma_{h\sigma}^{\ket{\psi}\leftarrow\ket{\psi'}}\rho_{\ket{\psi'}}-\Gamma_{h\sigma}^{\ket{\psi'}\leftarrow\ket{\psi}}\rho_{\ket{\psi}}\right),\label{eq:L_holes}\\
\bra{\psi}\mathcal{L}_{QP}\ket{\psi}&=\sum_{\psi'} \left(\Gamma_{QP}^{\ket{\psi}\leftarrow\ket{\psi'}}\rho_{\ket{\psi'}}-\Gamma_{QP}^{\ket{\psi'}\leftarrow\ket{\psi}}\rho_{\ket{\psi}}\right),\label{eq:L_QP}\\
\bra{\psi}\mathcal{L}_R\ket{\psi}&=\sum_{\psi'} \left(\Gamma_R^{\ket{\psi}\leftarrow\ket{\psi'}}\rho_{\ket{\psi'}}-\Gamma_R^{\ket{\psi'}\leftarrow\ket{\psi}}\rho_{\ket{\psi}}\right).\label{eq:L_relax}
\end{align}
The hole refilling rates according to Eq.~\eqref{eq:L_holes} are 
\begin{align}
\Gamma_{h\sigma}^{\ket{\psi}\leftarrow\ket{\psi'}}=\Gamma_{h}\left|\bra{\psi}h_\sigma^\dagger\ket{\psi'}\right|^2.
\end{align}
Since hole eigenstates are just given by the number states, the hole refilling rate can either be $\Gamma_{h}$ if refilling is possible or 0 otherwise. So, hole refilling does not depend on $\varepsilon_e$.
The QP rates in Eq.~\eqref{eq:L_QP} are given by
\begin{align}
\Gamma_{QP}^{\ket{\psi}\leftarrow\ket{\psi'}}=\Gamma_{QP}\left(\left|\bra{\psi}c\mathcal{P}\ket{\psi'}\right|^2+\left|\bra{\psi}c^\dagger \mathcal{P}\ket{\psi'}\right|^2\right).
\end{align}
For the eigenstates in Eq.~\eqref{eq:eigenstate_components}, we obtain explicitly
\begin{align}
\begin{split}
\Gamma_{QP}^{\ket{O_m}\leftarrow\ket{E_n}}=\Gamma_{QP} \big( &\left|\varphi_{2m}^{o*}\varphi_{2n}^{e}+\varphi_{3m}^{o*}\varphi_{3n}^{e} \right|^2\\
+ & \left|\varphi_{1m}^{o*}\varphi_{1n}^{e}+\varphi_{4m}^{o*}\varphi_{4n}^{e} \right|^2 \big),
\end{split}\\
\begin{split}
\Gamma_{QP}^{\ket{E_m}\leftarrow\ket{O_n}}
=\Gamma_{QP} \big( & \left|\varphi_{1m}^{e*}\varphi_{1n}^{o}+\varphi_{4m}^{e*}\varphi_{4n}^{o} \right|^2\\ +&\left|\varphi_{2m}^{e*}\varphi_{2n}^{o}+\varphi_{3m}^{e*}\varphi_{3n}^{o} \right|^2 \big).
\end{split}
\end{align}
The spin relaxation rates in Eq.~\eqref{eq:L_relax} are
\begin{align}
\Gamma_R^{\ket{\psi}\leftarrow\ket{\psi'}}=
\Gamma_R \left| \bra{\psi}d_\downarrow^\dagger d_\uparrow\ket{\psi'} \right| ^2 ,
\end{align}
and using the eigenstates in Eq.~\eqref{eq:eigenstate_components} explicitly
\begin{align}
\Gamma_R^{\ket{E_m}\leftarrow\ket{E_n}}=\Gamma_R \left| \varphi_{3m}^{e*} \varphi_{2n}^e \right|^2, \quad m\neq n,\\
\Gamma_R^{\ket{O_m}\leftarrow\ket{O_n}}=\Gamma_R \left| \varphi_{3m}^{o*} \varphi_{2n}^o \right|^2, \quad m\neq n.
\end{align}

Note that a spin relaxation process will always cause a transition to another eigenstate.
As we showed, QP and spin relaxation rates depend on $\varepsilon_e$ and require an explicit calculation via the components of the eigenstates. 
The full master equation is given by
\begin{widetext}
\begin{subequations}\label{eq:MG}
	\begin{align}
	\dot\rho_{\ket{E_n,\Uparrow\Downarrow}}(\tau)&=
	\sum_{m=1}^4 \left( \big(-W^L_{\ket{O_m}\leftarrow \ket{E_n}}-W^R_{\ket{O_m}\leftarrow \ket{E_n}}\big) \rho_{\ket{E_n,\Uparrow\Downarrow}}(\tau)+\Gamma_{QP}^{\ket{E_n} \leftarrow \ket{O_m}} \rho_{\ket{O_m,\Uparrow\Downarrow}}(\tau)-\Gamma_{QP}^{\ket{O_m} \leftarrow \ket{E_n}} \rho_{\ket{E_n,\Uparrow\Downarrow}}(\tau)  \right) \notag\\
	&+\sum_{m\neq n} \left(\Gamma_R^{\ket{E_n}\leftarrow \ket{E_m}}\rho_{\ket{E_m,\Uparrow\Downarrow}}(\tau)-\Gamma_R^{\ket{E_m}\leftarrow \ket{E_n}}  \rho_{\ket{E_n,\Uparrow\Downarrow}}(\tau)\right)	+\Gamma_{h\Uparrow}  \rho_{\ket{E_n,\Downarrow}}(\tau)+\Gamma_{h\Downarrow}  \rho_{\ket{E_n,\Uparrow}}(\tau),
	\\
	\dot\rho_{\ket{O_n,\Uparrow\Downarrow}}(\tau)&=
	\sum_{m=1}^4 \left( \big(-W^L_{\ket{E_m}\leftarrow \ket{O_n}}-W^R_{\ket{E_m}\leftarrow \ket{O_n}}\big) \rho_{\ket{O_n,\Uparrow\Downarrow}}(\tau)+\Gamma_{QP}^{\ket{O_n} \leftarrow \ket{E_m}} \rho_{\ket{E_m,\Uparrow\Downarrow}}(\tau)-\Gamma_{QP}^{\ket{E_m} \leftarrow \ket{O_n}} \rho_{\ket{O_n,\Uparrow\Downarrow}}(\tau)  \right)
	\notag\\
	&+\sum_{m\neq n} \left(\Gamma_R^{\ket{O_n}\leftarrow \ket{O_m}}\rho_{\ket{O_m,\Uparrow\Downarrow}}(\tau)-\Gamma_R^{\ket{O_m}\leftarrow \ket{O_n}}  \rho_{\ket{O_n,\Uparrow\Downarrow}}(\tau)\right)+\Gamma_{h\Uparrow}  \rho_{\ket{O_n,\Downarrow}}(\tau)+\Gamma_{h\Downarrow}  \rho_{\ket{O_n,\Uparrow}}(\tau),
	\\
	\dot\rho_{\ket{E_n,\Downarrow}}(\tau)&=
	\sum_{m=1}^4 \left(W^R_{\ket{E_n}\leftarrow \ket{O_m}} \rho_{\ket{O_m,\Uparrow\Downarrow}}(\tau)-W^L_{\ket{O_m}\leftarrow \ket{E_n}} \rho_{\ket{E_n,\Downarrow}}(\tau)+\Gamma_{QP}^{\ket{E_n} \leftarrow \ket{O_m}} \rho_{\ket{O_m,\Downarrow}}(\tau)-\Gamma_{QP}^{\ket{O_m} \leftarrow \ket{E_n}} \rho_{\ket{E_n,\Downarrow}}(\tau)  \right)\notag\\
	&+\sum_{m\neq n} \left(\Gamma_R^{\ket{E_n}\leftarrow \ket{E_m}}\rho_{\ket{E_m,\Downarrow}}(\tau)-\Gamma_R^{\ket{E_m}\leftarrow \ket{E_n}}  \rho_{\ket{E_n,\Downarrow}}(\tau)\right)	-\Gamma_{h\Uparrow}  \rho_{\ket{E_n,\Downarrow}}(\tau)+\Gamma_{h\Downarrow}  \rho_{\ket{E_n,0}}(\tau),
	\\	
	\dot\rho_{\ket{O_n,\Downarrow}}(\tau)&=
	\sum_{m=1}^4 \left(W^R_{\ket{O_n}\leftarrow \ket{E_m}} \rho_{\ket{E_m,\Uparrow\Downarrow}}(\tau)-W^L_{\ket{E_m}\leftarrow \ket{O_n}} \rho_{\ket{O_n,\Downarrow}}(\tau)+\Gamma_{QP}^{\ket{O_n} \leftarrow \ket{E_m}} \rho_{\ket{E_m,\Downarrow}}(\tau)-\Gamma_{QP}^{\ket{E_m} \leftarrow \ket{O_n}} \rho_{\ket{O_n,\Downarrow}}(\tau)  \right)\notag \\
	&+\sum_{m\neq n} \left(\Gamma_R^{\ket{O_n}\leftarrow \ket{O_m}}\rho_{\ket{O_m,\Downarrow}}(\tau)-\Gamma_R^{\ket{O_m}\leftarrow \ket{O_n}}  \rho_{\ket{O_n,\Downarrow}}(\tau)\right)-\Gamma_{h\Uparrow}  \rho_{\ket{O_n,\Downarrow}}(\tau)+\Gamma_{h\Downarrow}  \rho_{\ket{O_n,0}}(\tau),
	\\
	\dot\rho_{\ket{E_n,\Uparrow}}(\tau)&=
	\sum_{m=1}^4 \left(W^L_{\ket{E_n}\leftarrow \ket{O_m}} \rho_{\ket{O_m,\Uparrow\Downarrow}}(\tau)-W^R_{\ket{O_m}\leftarrow \ket{E_n}} \rho_{\ket{E_n,\Uparrow}}(\tau)+\Gamma_{QP}^{\ket{E_n} \leftarrow \ket{O_m}} \rho_{\ket{O_m,\Uparrow}}(\tau)-\Gamma_{QP}^{\ket{O_m} \leftarrow \ket{E_n}} \rho_{\ket{E_n,\Uparrow}}(\tau)  \right)\notag \\
	&+\sum_{m\neq n} \left(\Gamma_R^{\ket{E_n}\leftarrow \ket{E_m}}\rho_{\ket{E_m,\Uparrow}}(\tau)-\Gamma_R^{\ket{E_m}\leftarrow \ket{E_n}}  \rho_{\ket{E_n,\Uparrow}}(\tau)\right)-\Gamma_{h\Downarrow}  \rho_{\ket{E_n,\Uparrow}}(\tau)+\Gamma_{h\Uparrow}  \rho_{\ket{E_n,0}}(\tau),
	\\
	\dot\rho_{\ket{O_n,\Uparrow}}(\tau)&=
	\sum_{m=1}^4 \left(W^L_{\ket{O_n}\leftarrow \ket{E_m}} \rho_{\ket{E_m,\Uparrow\Downarrow}}(\tau)-W^R_{\ket{E_m}\leftarrow \ket{O_n}} \rho_{\ket{O_n,\Uparrow}}(\tau)+\Gamma_{QP}^{\ket{O_n} \leftarrow \ket{E_m}} \rho_{\ket{E_m,\Uparrow}}(\tau)-\Gamma_{QP}^{\ket{E_m} \leftarrow \ket{O_n}} \rho_{\ket{O_n,\Uparrow}}(\tau)  \right)\notag \\
	&+\sum_{m\neq n} \left(\Gamma_R^{\ket{O_n}\leftarrow \ket{O_m}}\rho_{\ket{O_m,\Uparrow}}(\tau)-\Gamma_R^{\ket{O_m}\leftarrow \ket{O_n}}  \rho_{\ket{O_n,\Uparrow}}(\tau)\right)-\Gamma_{h\Downarrow}  \rho_{\ket{O_n,\Uparrow}}(\tau)+\Gamma_{h\Uparrow}  \rho_{\ket{O_n,0}}(\tau),
	\\
	\dot\rho_{\ket{E_n,0}}(\tau)&=
	\sum_{m=1}^4\left(W^L_{\ket{E_n}\leftarrow \ket{O_m}} \rho_{\ket{O_m,\Downarrow}}(\tau)+W^R_{\ket{E_n}\leftarrow \ket{O_m}} \rho_{\ket{O_m,\Uparrow}}(\tau)+\Gamma_{QP}^{\ket{E_n} \leftarrow \ket{O_m}} \rho_{\ket{O_m,0}}(\tau)-\Gamma_{QP}^{\ket{O_m} \leftarrow \ket{E_n}} \rho_{\ket{E_n,0}}(\tau)  \right)\notag \\
	&+\sum_{m\neq n} \left(\Gamma_R^{\ket{E_n}\leftarrow \ket{E_m}}\rho_{\ket{E_m,0}}(\tau)-\Gamma_R^{\ket{E_m}\leftarrow \ket{E_n}}  \rho_{\ket{E_n,0}}(\tau)\right)-(\Gamma_{h\Downarrow}+\Gamma_{h\Uparrow})  \rho_{\ket{E_n,0}}(\tau),
	\end{align}
	\begin{align}
	\dot\rho_{\ket{O_n,0}}(\tau)&=
	\sum_{m=1}^4 \left(W^L_{\ket{O_n}\leftarrow \ket{E_m}} \rho_{\ket{E_m,\Downarrow}}(\tau)+W^R_{\ket{O_n}\leftarrow \ket{E_m}} \rho_{\ket{E_m,\Uparrow}}(\tau)+\Gamma_{QP}^{\ket{O_n} \leftarrow \ket{E_m}} \rho_{\ket{E_m,0}}(\tau)-\Gamma_{QP}^{\ket{E_m} \leftarrow \ket{O_n}} \rho_{\ket{O_n,0}}(\tau)  \right)\notag \\
	&+\sum_{m\neq n} \left(\Gamma_R^{\ket{O_n}\leftarrow \ket{O_m}}\rho_{\ket{O_m,0}}(\tau)-\Gamma_R^{\ket{O_m}\leftarrow \ket{O_n}}  \rho_{\ket{O_n,0}}(\tau)\right)
	-(\Gamma_{h\Downarrow}+\Gamma_{h\Uparrow})  \rho_{\ket{O_n,0}}(\tau),
	\end{align}
\end{subequations}
\end{widetext}
for $n=1,2,3,4$. The stationary state is given by $\partial_\tau \rho^{\rm stat}_{\ket{\psi}}(\tau)=0.$

\section{Fano factor and $g^{(2)}$ correlation function}\label{app:fano}
\begin{figure}[t]
	\centering
	\includegraphics[width=\columnwidth]{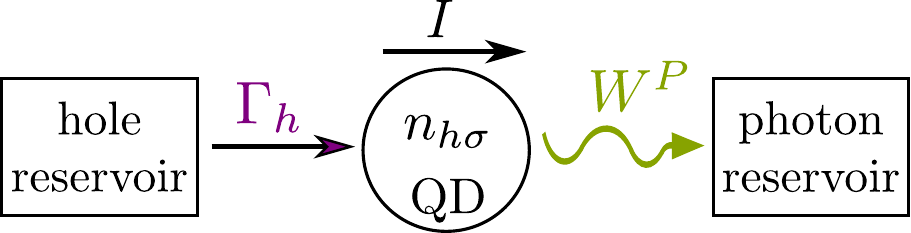}
	\caption{Transitions between holes on the QD and the coupled reservoirs. Holes tunnel onto the QD from the hole reservoir via $\Gamma_{h}$ and recombine with electrons to photons via photon emission with rates $W^P$. The calculated current $I$ flows from the left to the right.}
	\label{fig:hole_current}
\end{figure}

To calculate the Fano factor \cite{Fano1947}, we use a generalized master equation approach \cite{Flindt_2004,Kaiser2007,Dominguez2010} derived for charge transport through QDs. The Fano factor is defined as $F=S/e I$ with the elementary charge $e$, the average current $I$ and the zero-frequency current noise $S$. $F$ gives a measure of whether a system compared to a Poisson process ($F=1$) is in the sub-Poissonian ($F<1$) or super-Poissonian regime ($F>1$). 
In our setup, the holes flow from the hole reservoir onto the QD via the hole refilling rate $\Gamma_{h}$ and recombine with an electron to a photon by recombination rates $W^P_{\ket{\psi_f}\leftarrow\ket{\psi_i}}$ defined in Eq.~\eqref{eq:changerate}. In the stationary limit, the number of holes tunneled from the hole reservoir onto the QD is equal to the number of holes that left the QD via photon emission (i.e. $(d/dt) \langle n_{h\sigma}\rangle=0$). Since each emitted photon also annihilates an electron in the QD, the average current of holes tunneling onto the QD equals that of the electrons leaving the QD (with opposite sign) so that a current $I$ through the QD can be defined, see Fig.~\ref{fig:hole_current}. In the following, we focus on the contribution to the current where holes leave the QD via photon emission. In this sense, the charge current $I$ can be understood as an emission intensity of photons with $\sum_PI_P=I/e$, introduced in Eq.~\eqref{eq:total_int}.

We use the approach and notation according to \cite{Flindt_2004}. To find the contributions to the current of interest, we decompose the Liouvillian as $\mathcal{L}=\mathcal{L}_0+\mathcal{J}$, where the jump superoperator $\mathcal{J}$ describes incoherent transitions between the system and the photon reservoir, whereas $\mathcal{L}_0$ includes all remaining processes. The Liouvillian for the full model can be deduced from the full master equation in Eq.~\eqref{eq:MG} according to $\dot\rho(\tau)=\mathcal{L}\rho(\tau)$, where $\rho(\tau)$ is the vector of all matrix elements $\rho_{\ket{\psi}}(\tau)$ of the reduced density matrix.
Therefore, the master equation can be written as
\begin{align}\label{eq:ME_n}
\dot\rho^{(n)}(\tau)=\mathcal{L}_0\rho^{(n)}(\tau)+\mathcal{J}\rho^{(n-1)}(\tau),
\end{align}
where $\rho^{(n)}$ denotes the number resolved density matrix with $n$ the number of holes that recombined with electrons to photons, i.e. the number of emitted photons.
We identify the terms in the master equation that are responsible for these recombination processes and show explicitly the current superoperator and the density operator
\begin{widetext}
\begin{align}
\mathcal{J}=
\begin{pmatrix}
0 &
0 &
W^P_{\ket{E_{\sigma+}}\leftarrow \ket{O_{\sigma+}}} &
W^P_{\ket{E_{\sigma+}}\leftarrow \ket{O_{\sigma-}}} \\
0 &
0 &
W^P_{\ket{E_{\sigma-}}\leftarrow \ket{O_{\sigma+}}} &
W^P_{\ket{E_{\sigma-}}\leftarrow \ket{O_{\sigma-}}} \\
W^P_{\ket{O_{\sigma+}}\leftarrow \ket{E_{\sigma+}}} &
W^P_{\ket{O_{\sigma+}}\leftarrow \ket{E_{\sigma-}}} &
0 & 0 \\
W^P_{\ket{O_{\sigma-}}\leftarrow \ket{E_{\sigma+}}} &
W^P_{\ket{O_{\sigma-}}\leftarrow \ket{E_{\sigma-}}} &
0 & 0  \\
\end{pmatrix}, \quad
\rho(\tau)=
\begin{pmatrix}
\rho_{\ket{E_{\sigma+}}}(\tau)\\
\rho_{\ket{E_{\sigma-}}}(\tau)\\
\rho_{\ket{O_{\sigma+}}}(\tau)\\
\rho_{\ket{O_{\sigma-}}}(\tau)
\end{pmatrix},
\end{align}
\end{widetext}
respectively, for the spinless model in Eq.~\eqref{eq:mastereq_spinless}.
The current superoperator $\mathcal{J}$ for the full model can be constructed analogously. Note that for the full model, where $\Gamma_{h}$ is finite, we could also define the average current as the holes that tunneled from the hole reservoir onto the QD via $\Gamma_{h}$.

From the $n$-resolved density operator in Eq.~\eqref{eq:ME_n} one can obtain the complete probability distribution for $n$ holes that recombined with electrons to photons. The probability distribution gives access to the cumulant generating function. From this one can obtain the first two current cumulants which are the current and the zero-frequency noise.

Following the notation in \cite{Flindt_2004} taking the trace $\rm Tr(\mathcal{L}\rho^{\rm stat})=0$ corresponds to multiplying $\mathcal{L}$ from the right with $\ket{0}\rangle$ which is the vector of the stationary state occupations $\rho^{\rm stat}_{\ket{\psi}}$ and from the left with a vector $\langle\bra{\tilde0}$ where every element is equal to one. With this, one can introduce the projectors $\mathcal{P}_0=\ket{0}\rangle\langle\bra{\tilde 0}$ and $\mathcal{Q}=1-\mathcal{P}_0$ with the properties $\mathcal{P}_0\mathcal{L}=\mathcal{LP}_0=0$ and $\mathcal{QLQ}=\mathcal{L}$. One can define the pseudoinverse of the Liouvillian $\mathcal{R}=\mathcal{QL}^{-1}\mathcal{Q}$, which is needed for calculating the second cumulant. Here, the inversion of the singular matrix $\mathcal{L}$ is only executed in the $\mathcal{Q}$ subspace, where $\mathcal{L}$ is regular. With this, one obtains for the average current
\begin{align}
I=e\braket{\bra{\tilde 0}\mathcal{J}\ket{0}},
\end{align}
with $e>0$ and for the zero-frequency current noise
\begin{align}
S=e^2\braket{\bra{\tilde 0}\mathcal{J}-2 \mathcal{J} \mathcal{R} \mathcal{J}\ket{0}}.
\end{align}

Since from the Fano factor one cannot unambiguously determine whether the emitted photons are bunched or antibunched, we also calculate the second-order correlation function $g^{(2)}$ which measures the correlation between two system jumps separated by a time $\tau$. The correlation function can be defined with the current superoperator as \cite{Emary_2012}
\begin{align}
g^{(2)}(\tau)=\frac{\braket{\bra{\tilde 0}\mathcal{J}\Omega(\tau)\mathcal{J}\ket{0}}}{\braket{\bra{\tilde 0}\mathcal{J}\ket{0}}^2},
\end{align}
where $\Omega(\tau)=\exp(\mathcal{L}\tau)$ is the the master equation propagator.
\begin{figure}[t]
	\includegraphics[width=\columnwidth]{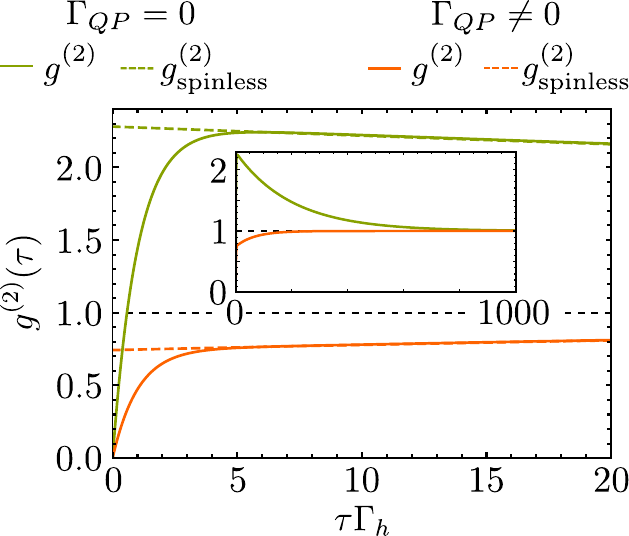}
	\caption{Correlation function $g^{(2)}(\tau)$. We show $g^{(2)}(\tau)$ over $\tau$ for the QD energy $\varepsilon_e=\Delta_{Z,e}-\xi$ for the full model (solid lines) and $g_{\rm spinless}^{(2)}(\tau)$ over $\tau$ according to Eq.~\eqref{eq:g2} for the spinless model (dashed lines). The inset shows longer times $\tau$ where $g^{(2)}(\tau)\to1$. For the full model, we used $\xi=0.04\Delta_{Z,e}$, $10t_{1\uparrow}=t_{1\downarrow}=0.025\Delta_{Z,e}$ $t_2=0$, $\Gamma_{h}=100 W_{\rm max}$, $\Gamma_R=0$, $\Gamma_{QP}=0$ (green) and $\Gamma_{QP}=W_{\rm max}$ (orange). For the spinless model, we used $\xi=0.04\Delta_{Z,e}$, $t_{1\sigma}=0.025\Delta_{Z,e}$, $\Gamma_{QP}=0$ (green) and $\Gamma_{QP}=W_{\rm max}$ (orange).}
	\label{fig:g2}
\end{figure}
By comparing the $g^{(2)}$ functions at two different times separated by $\tau$, one can identify
\begin{equation}\label{eq:g2comp}
\begin{array}{ll}
g^{(2)}(0)>g^{(2)}(\tau) \quad& \text{bunching},\\
g^{(2)}(0)<g^{(2)}(\tau) \quad& \text{antibunching}.
\end{array}
\end{equation}
Especially, the $g^{(2)}$ function is also related to the Fano factor by  \cite{Emary_2012}
\begin{align}
F=1+\frac{2 I}{e} \int_{0}^\infty d \tau (g^{(2)}(\tau)-1).
\end{align}
So if the integral over $g^{(2)}(\tau)-1$ is positive (negative) the corresponding Fano factor is super-Poissonian (sub-Poissonian).

For the spinless model (where we set $\Gamma_{h}\to\infty$), we can solve the model analytically. We give the analytical expressions for $g^{(2)}$ and the Fano factor for $t_2=0$ at $\varepsilon_{e\sigma}=\pm\xi$,
\begin{widetext}
\begin{align}
&g_{\rm spinless}^{(2)}(\tau)=1+e^{-\left(\tilde\Gamma_{QP}+\frac{1}{2}\right) \tau W_{\rm{max}} }
\underbrace{\frac{ \left(\left(1-4 \tilde\Gamma_{QP}^2\right) \xi ^2 t_{1\sigma}^2-4 \tilde\Gamma_{QP}^2 \xi ^4\right)}{2 \left(2 \tilde\Gamma_{QP} \xi ^2+(2 \tilde\Gamma_{QP}+1) t_{1\sigma}^2\right)^2}}_{c_0},\label{eq:g2}\\
&F_{\rm spinless}= 1-\frac{2 \tilde\Gamma_{QP}}{(2 \tilde\Gamma_{QP}+1)^2}
+t_{1\sigma}^2 \left(\frac{1}{2 \tilde\Gamma_{QP} \xi ^2+2 \tilde\Gamma_{QP} t_{1\sigma}^2+t_{1\sigma}^2}
-\frac{1}{(2 \tilde\Gamma_{QP}+1)^2 \left(\xi ^2+t_{1\sigma}^2\right)}\right),\label{eq:Fano}
\end{align}
\end{widetext}
with $\tilde{\Gamma}_{QP}=\Gamma_{QP}/W_{\rm max}$.
Therefore, emitted photons are, according to Eq.~\eqref{eq:g2comp},
\begin{equation}\label{eq:QPcrit}
\begin{array}{ll}
\text{bunched,} & \quad \text{if } \tilde\Gamma_{QP}<\frac{|t_{1\sigma}|}{2 \sqrt{\xi ^2+t_{1\sigma}^2}},\\
\text{antibunched,} & \quad \text{if } \tilde\Gamma_{QP}>\frac{|t_{1\sigma}|}{2 \sqrt{\xi ^2+t_{1\sigma}^2}}.
\end{array}
\end{equation}
In the case of no QP, Eq.~\eqref{eq:g2} and Eq.~\eqref{eq:Fano} simplify to
\begin{align}
&g_{\rm spinless}^{(2)}(\tau)=1+e^{-\frac{\tau W_{\rm max}}{2}}\frac{\xi ^2 }{2 t_{1\sigma}^2},\\
&F_{\rm spinless}=2-\frac{t_{1\sigma}^2}{\xi ^2+t_{1\sigma}^2}=2-\frac{2}{e} I_{\rm spinless}.
\end{align}
Interestingly, the Fano factor can be written using the average current $I_{\rm spinless}$. Note that for $\xi=0$, where even and odd states are degenerate, at resonance $\varepsilon_{e\sigma}=0$ the Fano factor $F_{\rm spinless}=1$ and $g^{(2)}_{\rm spinless}(\tau)=1,\forall\tau$. Here, the system is Poissonian and the emitted photons are completely uncorrelated.

In Fig.~\ref{fig:g2}, we show the $g^{(2)}$ function of the full model (solid lines) at $\varepsilon_e=\Delta_{Z,e}-\xi$ in comparison to the spinless model (dashed lines) without and with QP. For the full model, the $g^{(2)}$ function starts at around 0 and increases rapidly. After a time of the order of $\tau\sim1/\Gamma_h$ the curves flatten and follow $g^{(2)}_{\rm spinless}$. Here,  $g^{(2)}(0)<g^{(2)}(\tau), \forall{\tau}$, hence emitted photons are antibunched, according to Eq.~\eqref{eq:g2comp}.

The difference between the full and the spinless model can be explained by the time that holes need to get refilled on the QD, since in the limit $\Gamma_h\to\infty$ for the spinless model the refilling time is zero. If we fit the $g^{(2)}$ function \cite{Kurtsiefer2000} for the full model ($\Gamma_h>\Gamma_{QP},W_{\rm max}$),
\begin{align}\label{eq:g2fit}
g^{(2)}(\tau)\approx\underbrace{1+ c_0 e^{-(\tilde{\Gamma}_{QP}+\frac{1}{2})\tau W_{\rm max}}}_{g^{(2)}_{\rm spinless}(\tau)}-(1+c_0) e^{-\Gamma_h \tau},
\end{align}
we find two exponential functions with different decay times. Since emitting one photon is equivalent to an electron and a hole that leave the QD, the $g^{(2)}$ function mirrors what happens in the hole and electronic subsystem. Here, the dynamics at short times $\tau\sim\frac{1}{\Gamma_h}$ and at later times $\tau\sim\frac{1}{\Gamma_{QP}+\frac{W_{\rm max}}{2}}$ compete \cite{Emary_2012}. At short times, we can identify that the dynamics in the hole subsystem dominates, where hole refilling is antibunched due to fermionic statistics. After refilling, the holes are mostly in the doubly occupied state before the next photon is emitted and so at later times the dynamics of the system is governed by the dynamics in the electronic subsystem.
\begin{figure}[t!]
	\includegraphics[width=.85\columnwidth]{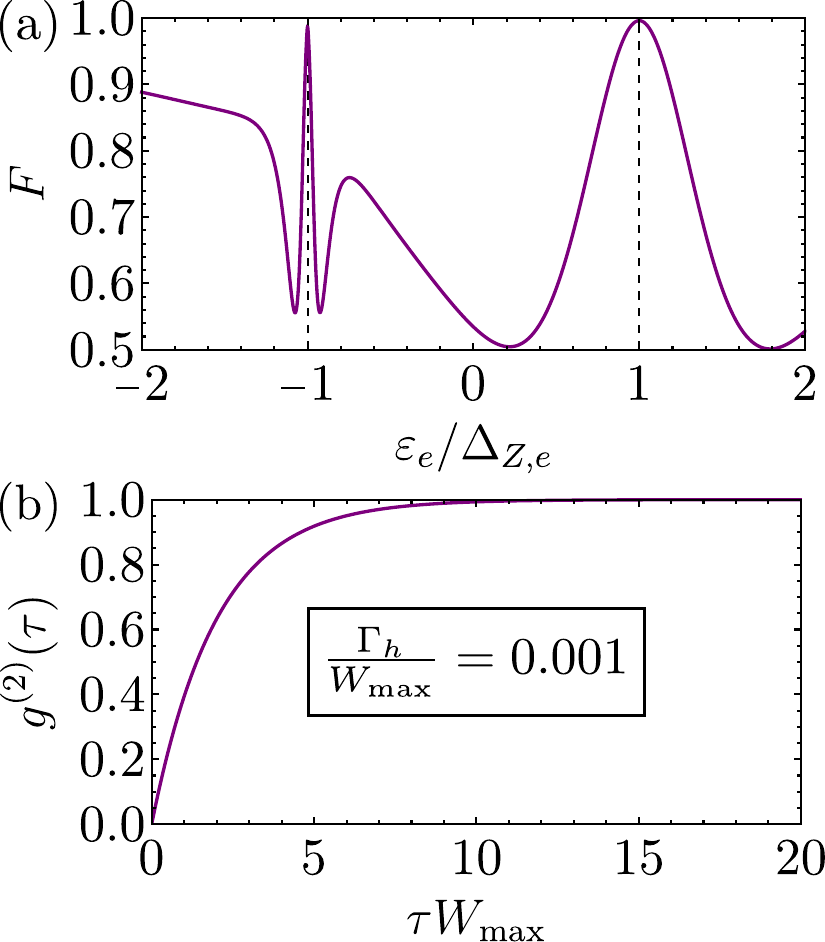}
	\caption{Fano factor and correlation function $g^{(2)}(\tau)$ for slow holes. We show the Fano factor $F$ over the QD energy $\varepsilon_e$ in (a) and $g^{(2)}(\tau)$ over $\tau$ for $\varepsilon_e=\Delta_{Z,e}$ in (b). For both plots we used $\xi=0$, $10t_{1\uparrow}=t_{1\downarrow}=0.025\Delta_{Z,e}$, $t_2=0$, $\Gamma_{h}=0.001 W_{\rm max}$, $\Gamma_R=0$, $\Gamma_{QP}=0$.}
	\label{fig:fano_slow_holes}
\end{figure}

In the ideal case ($\Gamma_{QP}=0$) and after short times, $g^{(2)}(\tau)$ becomes much larger than 1 and has a negative slope, see Fig.~\ref{fig:g2} (green, solid). Here, electrons on the QD are bunched and thus emitted photons show bunching on timescales $\tau\sim\frac{2}{W_{\rm max}}$. For ${\tilde \Gamma}_{QP}$ larger than the critical value defined in Eq.~\eqref{eq:QPcrit}, $g^{(2)}(\tau)\leq1$ and has a positive slope, see Fig.~\ref{fig:g2} (orange, solid), thus emitted photons are antibunched. Here, we can also identify two processes that happen on different timescales. The antibunching effect in the electronic subsystem happens on a larger timescale $\tau\sim\frac{1}{\Gamma_{QP}+\frac{W_{\rm max}}{2}}$ than in the hole subsystem.

The comparison between the spinless and the full model shows that in the electronic subsystem without QP there exists a process that causes bunching of photons leading to a super-Poissonian Fano factor, see Fig. \ref{fig:emission_cycle_fano}(b) (green).

For slow holes ($\Gamma_h<W_{\rm max}$) and decoupled spins at $\varepsilon_e=\pm\Delta_{Z,e}$ with $\xi=0$ and $t_2=0$, we can deduce from a fit of the full model that $g^{(2)}(\tau)\approx 1-e^{-\left(\frac{W_{\rm max}}{2}+\Gamma_h\right)\tau}$ like for a single resonant level between two reservoirs \cite{Emary_2012}. Consequently,  $F$ can be calculated with the current formula $I=e\frac{W_{\rm max}\Gamma_h}{W_{\rm max}+2\Gamma_h}$ as $F=1-\frac{\Gamma_h W_{\rm max}}{\left(\Gamma_h+\frac{W_{\rm max}}{2}\right)^2}$ near the resonances between electrons and MBSs and thus the system processes are sub-Poissonian and emitted photons are always antibunched. In Fig.~\ref{fig:fano_slow_holes}, we show the Fano factor and the $g^{(2)}$ function for slow holes (and $\xi=0$) where $\Gamma_h\ll W_{\rm{max}}$ and the noise is dominated by the hole refilling with $F\leq 1$.

\section{Total intensity including QP}\label{app:int_QP}
The coefficients in $I_\sigma$ in Eq.~\eqref{eq:int_QP} are given by
\begin{subequations}
\begin{align}
&D_1=\frac{2(2 \tilde\Gamma_{QP}+1) t_{1\sigma}^2 \left(\tilde\Gamma_{QP} \xi ^2+(2 \tilde\Gamma_{QP}+1) t_{1\sigma}^2\right)}{\tilde\Gamma_{QP}},\end{align}\begin{align}
&D_2=2(2 \tilde\Gamma_{QP}+1) t_{1\sigma}^2,\end{align}\begin{align}
&D_3=-\frac{(2 \tilde\Gamma_{QP} t_{1\sigma}+t_{1\sigma})^2 \left((1-2 \tilde\Gamma_{QP})^2 t_{1\sigma}^2-8 \tilde\Gamma_{QP} \xi ^2\right)}{4 \tilde\Gamma_{QP}^2},\end{align}\begin{align}
&D_4^2=\xi ^2-\frac{(2 \tilde\Gamma_{QP} t_{1\sigma}+t_{1\sigma})^2}{2 \tilde\Gamma_{QP}},
\end{align}
\end{subequations}
where $\tilde\Gamma_{QP}=\Gamma_{QP}/W_{\rm max}$ is a dimensionless rate.

\end{document}